\documentclass[aps,prb,twocolumn,float,amsmath]{revtex4}
\usepackage[dvips]{color}\usepackage{hyperref}
\usepackage[normalem]{ulem} 

\definecolor{g-blue}{rgb}{0.83,0.95,1}
\definecolor{g-yellow}{rgb}{1,1,0.7}
\definecolor{g-green}{rgb}{0.9,1,0.9}
\definecolor{green}{rgb}{0,0.6,0}
\definecolor{cyan}{rgb}{0,0.7,0.7}
\definecolor{black}{rgb}{0,0,0}
\definecolor{grey}{rgb}{0.4 ,0.4 ,0.4 }

\def\white#1{\textcolor{white}{#1}}
\usepackage{bm}
\usepackage{amssymb}
\usepackage{amsfonts}
\usepackage{amsmath}
 \usepackage{graphicx}
  \usepackage{amsmath,bm,epsfig}

\def \ed {\end{document}}
\def\Fbox#1{\vskip1ex\hbox to 8.5cm{\hfil\fboxsep0.3cm\fbox{%
  \parbox{8.0cm}{#1}}\hfil}\vskip1ex\noindent}  

\newcommand{\Eq}[1]{Eq.\,(\ref{#1})}
\newcommand{\Eqs}[1]{Eqs.\,(\ref{#1})}
\newcommand{\Fig}[1]{Fig.\,\ref{#1}}
\newcommand{\Figs}[1]{Figs.\,\ref{#1}}
\newcommand{\Sec}[1]{Sec.\,\ref{#1}}
\newcommand{\Ref}[1]{Ref.\,\cite{#1}}
\newcommand{\Refs}[1]{Refs.\,\cite{#1}}

\def\be{\begin{equation}}\def\ee{\end{equation}}
\def\bea{\begin{eqnarray}}\def\eea{\end{eqnarray}}
\def\bse{\begin{subequations}}\def\ese{\end{subequations}}

\let \nn  \nonumber

\let \= \equiv \let\*\cdot \let\~\widetilde \let\^\widehat \let\-\overline

  \def\1{\bm1} 

\def\<{\left\langle}    \def\>{\right\rangle}
\def\({\left(}          \def\){\right)}
 \def \[ {\left [} \def \] {\right ]}

\newcommand{\ve}{\varepsilon}

\newcommand{\B}[1]{{\bm{#1}}}
\newcommand{\C}[1]{{\mathcal{#1}}}    

\renewcommand{\sb}[1]{_{\text {#1}}}  
\renewcommand{\sp}[1]{^{\text {#1}}}  
\newcommand{\Sp}[1]{^{^{\text {#1}}}} 
\def\Sb#1{_{\scriptscriptstyle\rm{#1}}}
\def\He4 {$^4$He~}

\begin{document}

\title{ Statistics of turbulence and intermittency enhancement in  superfluid $^4$He counterflow }

\author{   S. Bao${^{1,2}}$,         W. Guo${^{1,2}}$,  V. S. L'vov${^{3}}$,
 A. Pomyalov${^{3}}$ }

\affiliation{${^{1}}$ National High Magnetic Field Laboratory,  1800 East Paul Dirac Drive. Tallahassee,  FL 32310, USA,\\
${^{2}}$ Mechanical Engineering Department,  Florida State University,  Tallahassee, FL32310, USA,\\
${^{3}}$ Department of Chemical and Biological  Physics,  Weizmann Institute of Science,  Rehovot 76100,  Israel}

\begin{abstract} We report a detailed analysis of  the energy spectra, second- and high-order  structure functions of velocity differences in  superfluid $^4$He counterflow turbulence, measured in a wide range of temperatures and heat fluxes. We show  that the one-dimensional energy spectrum $E_{xz} (k_y)$ (averaged over the $xz$-plane, parallel to the channel wall),  directly measured  as a function of the  wall-normal wave-vector $k_y$,  gives more  detailed information on the energy distribution over scales than the corresponding second-order structure function $S_{2}(\delta_y)$. In particular,  we discover two intervals of $k_y$ with different apparent exponents: $E_{xz} (k_y)\propto k_y^{-m\Sb C}$ for $k\lesssim k_\times$ and  $E_{xz} (k_y)\propto k_y^{-m\Sb F}$ for $k\gtrsim k_\times$. Here $k_\times$ denotes wavenumber  that separate scales with relatively strong (for $k\lesssim k_\times$) and relatively weak (for $k\gtrsim k_\times$)  coupling between the normal-fluid and superfluid velocity components.   We interpret these $k$-ranges as \textit{cascade-dominated} and \textit{mutual friction-dominated} intervals, respectively. General behavior  of the experimental spectra $E_{xz}(k_y)$ agree well with the predicted spectra [Phys. Rev. B \textbf{97},  214513 (2018)]. Analysis of the $n$-th order structure functions statistics shows that in the energy-containing interval the statistics of counterflow turbulence is close to Gaussian, similar to  the classical hydrodynamic turbulence. In the  cascade- and mutual friction-dominated intervals we found some modest  enhancement of intermittency with respect of its level in classical turbulence. However, at  small  scales,
the  intermittency becomes much stronger than in the classical turbulence.

\end{abstract}
\maketitle
  \section{\label{s:1}Introduction}

Below the Bose-Einstein condensation temperature $T_\lambda\approx 2.17\,$K, liquid \He4 becomes a quantum   superfluid\,\cite{Donnely,DB98,2} with
the vorticity  constrained to vortex-line singularities of  core radius $a_0\approx  10^{-8}\,$cm and  fixed circulation $\kappa= h/M$~\ \cite{Feynman}.  Here $h$ is Planck's constant  and $M$ is the mass of the \He4 atom. The  superfluid turbulence is manifested  as a complex tangle of these vortex lines   with a typical inter-vortex distance  $\ell\sim 10^{-4}- 10^{-2}\,$cm\,~\cite{Vinen}.

Large-scale hydrodynamics of such system    is usually described by a  two-fluid model, interpreting $^4$He as a mixture of two coupled fluid components: an inviscid  superfluid and a viscous normal fluid.  The temperature dependent densities of the components $\rho\sb s(T)$ and  $\rho\sb n(T)$ define their contributions to the mixture. The total density of $^4$He $\rho=\rho\sb s(T)+ \rho\sb n(T)=\rho(T)$  weakly depends on the temperature.  The tangle of quantum vortexes mediates the interaction between fluid components  via mutual friction force \,\cite{Donnely,Vinen,Vinen2,Vinen3,37}.

There is  a building evidence  \cite{SS-2012,TenChapters,BLR, Roche-new} that the large-scale turbulence in mechanically driven superfluid \He4  is similar to the classical turbulence. In this case both components move with close velocities being coupled by the mutual friction force  almost at all scales.
	On the contrary, the turbulence generated in superfluid \He4  by a heat flux in a channel has no classical analogy. Here two components are moving  in the opposite directions relative to the channel walls, with respective mean velocities $\B U \sb n$ and  $\B U\sb s$.
	In this way  the counterflow velocity $\B U\sb {ns}=\B U \sb n - \B U\sb s\ne 0$, proportional to the applied heat flux, is created along the channel,  which can trigger the creation of a tangle of vortex lines above a small critical velocity.
	
	Systematic experimental studies of counterflow turbulence,  pioneered by classical 1957-papers   of Vinen\cite{Vinen3},  were long concentrated  mostly on global characteristics of the  vortex tangle, cf. see \Ref{SS-2012} for a review. The statistics of turbulent fluctuations was not accessible.  Recently, the turbulent statistics in the $^4$He normal component was measured in the form of the cross-stream 2$\sp {nd}$-order structure functions\,\cite{WG-2015,WG-2017}
	\begin{subequations}
\label{Sn}
 \begin{equation}\label{Sna}
	S_2 (Y)= \< |\Delta   u_x(Y,y,t)|^2 \> \ .
\end{equation}
Here $\< \dots \>$ denotes an ensemble average over many trials and $\Delta u_x(Y,y,t)$ is the
  streamwise velocity difference
\begin{equation}\label{Snb}
\Delta  u_x(Y,y,t) =u_x(y+Y,t)-u_x(y,t)\ .
\end{equation}
Other studies \cite{Prague1,Prague2} measured the statistics of the superfluid component.

	Recent theoretical analysis\,\cite{LP-2018} found that  the energy spectra in counterflow turbulence are not scale-invariant  and cannot be rigorously connected with apparent scaling exponents of the second-order velocity structure functions  measured  in \Ref{WG-2017} at modest values of the Reynolds numbers.   In this paper, we suggest  a new way to analyze the visualization data\,\cite{WG-2017}   that   allows  the one-dimensional energy spectra to be determined so that a direct comparison with theoretical predictions can be made\,\cite{LP-2018}. In addition, we use  higher-order structure functions to assess the level of intermittency in counterflow turbulence.

The paper is organized as follows:\\
 The analytical background of the problem of statistical description of superfluid counterflow turbulence is covered in Section\,\ref{s:2}. It starts with \Sec{ss:stat} which is devoted to the second-order statistical characteristics of homogeneous turbulence. In \Sec{ss:new}, we suggest a new way of velocity data analysis that allows to directly extract the one-dimensional energy spectra. The recent analytic theory of counterflow turbulence\cite{LP-2018}, required for our current  analysis, is summarized in \Sec{ss:theory}.
  The main result of the theory is the energy-balance \Eq{balance} that allows to find  the normal-fluid and superfluid  energy spectra in a wide range of the problem parameters.

  Our experimental results on  the  statistics of the normal-fluid turbulence and their analysis are presented in Section\,\ref{s:analysis}. In   \Sec{ss:exp}, we  briefly   describe the experimental techniques.  The important  cross-over wavenumbers for the current problem are estimated in \Sec{ss:est}.   Section\,\ref{ss:E2} is devoted to  the second-order statistics of counterflow turbulence: the velocity structure functions, $S_2(\delta)$ (see  \Sec{sss:S2} and left column of \Fig{F:1}), and  the energy spectra, $E(k)$ (see \Sec{sss:E} and   middle  column  of \Fig{F:1}). In particular,  we  demonstrate that the counterflow energy spectra can be divided in two sub-intervals: a cascade dominated interval and a mutual-friction dominated interval, with the apparent scaling exponents $m\Sb C \simeq 2$ and  $m\Sb F \simeq 3$ (see \Fig{F:1}, right column).   An important question about the relationship between $S_2(\delta)$ and $E(k)$ is discussed in \Sec{sss:viscosity} and illustrated in \Fig{F:2}. We show in  \Sec{ss:comp} and \Fig{F:3} that  the theoretically predicted energy spectra are in good agreement with the experimental energy spectra in the cascade-dominated range of wavenumbers.

In Section\,\ref{ss:flat}, we concentrate on high-order velocity structure functions:
\begin{equation}\label{Snc}
 S_4(Y)=\< |\Delta   u_x(Y,y,t)|^4 \>\,,\   S_6(Y)=\< |\Delta   u_x(Y,y,t)|^6 \>\ .
\end{equation}
\end{subequations}
In \Fig{F:4} we show that the flatness
 $ F_4(Y)$    and    hyper-flatness $F_6(Y)$,
 \begin{equation}\label{Snc}
 F_4(Y)=S_4(Y)/S_2^2(Y)\,,\quad   F_6(Y)=S_6(Y)/S_2^3(Y) \,,
 \end{equation}
have two ranges of power-law behavior with an apparent scaling of $F_4(Y)\propto Y^{x^{(1),(2)}_4}$ and  $F_6(Y)=S_6(Y)/S_2^3(Y)\propto Y^{x^{(1),(2)}_6}$, respectively.
For $Y$ larger than some $Y_*$ that corresponds to the cascade- and mutual friction-dominated subintervals of the energy spectra,   $x^{(1)}_4\simeq 0.20$ and $x^{(1)}_6\simeq 0.50$,   which are moderately larger than the inertial range exponents in classical hydrodynamic turbulence, i.e., $x_4\Sp{HT}\simeq 0.14$ and $x_6\Sp{HT}\simeq 0.38$. However, as $Y$ approaches the viscous scales (i.e., $Y\lesssim Y_*$), the small-scale intermittency becomes stronger:   $x^{(2)}_4\simeq 0.5$ and $x^{(2)}_6\simeq 1.4$. This behavior is similar to the intermitency enhancement observed in the mechanically driven \He4 \cite{He4-DNS,He4-coflow}.

 Final \Sec{s:sum}  briefly  summarizes our findings.

 \section{\label{s:2}Analytical background}

 \subsection{\label{ss:stat}Second-order statistical characteristics of homogeneous superfluid turbulence }
 The most general  statistical description of the homogeneous superfluid \He4 turbulent velocity field $\B u_j(\B r)$ at the level of the second-order statistics can be done in the terms of the three-dimensional (3D) cross-correlation functions of the normal-fluid and superfluid turbulent velocity fluctuations in the $\B k$-representation:
 \begin{subequations}\label{def-F}
\begin{equation}\label{def-Fa}
  (2\pi)^3 \delta (\B k + \B k')F^{\alpha\beta}  _{ij} (\bm k)=\<  v _i^\alpha(\bm k)  v^\beta _j(\bm k')\>\ .
 \end{equation}
Here  $\B v_j(\B k)$ is the Fourier transform  of  $\B u_j(\B r)$:
 \begin{equation}\label{def-Fb}
\B v_i(\B k)= \int \B u_i(\B r) \exp (i \B k \cdot \B r)\,  d \B r\,,
\end{equation}
$F_j(\bm k)\=   F^{\alpha\alpha}_{j}(\bm k)$,  $\alpha, \beta= \{x,y,z\}$ are vector indexes, subscripts ``$_{i,j}$"  denote  the normal-fluid
 (${i,j}=$n)    or  the superfuid (${i,j}=$s) components, and  $\B k$ is the  wave-vector. The inverse Fourier transform is defined as follows:
      \begin{equation}\label{def-Fc}
      \B u_i(\B r)= \int \B v_i(\B k) \exp (-i \B k \cdot \B r)\,   \frac{d\B k}{(2\pi)^3}\ .
      \end{equation}
 \end{subequations}

 The visualization technique,  to be discussed in more details in  \Sec{ss:exp},  allows one to measure the streamwise normal-fluid velocity across a channel, $v\sb{n}^x(y,t)$.
Henceforth,  unless stated explicitly, we consider only
 this velocity component, i.e    $i,j=$n, $\alpha=\beta=x$ and omit these indexes. For example, $F^{\alpha\beta}  _{ij} (\bm k) \Rightarrow F^{xx} \sb{nn} (\bm k)  \Rightarrow F  (\bm k)$.

 More compact, but less detailed information on the  statistics of turbulence is given by one-dimensional (1D) energy spectra  $E (k)$   averaged over all directions of vector $\B k$:
 \begin{subequations} \begin{equation}  \label{EspA}
 	E\sb{sp}  (k)=\frac{k^2}{(2\pi)^3} \int F (\B k) d\cos \theta\,  d \varphi\, ,
 	\end{equation}
 	Here we used spherical coordinates, with polar and azimuth angles $\theta$ and $\varphi$.  The polar angle is measured from the direction of   $\B U\sb {ns}$  .
 	
 	In the isotropic case $  F (\B k)=F (k)$, i.e., is independent of  $\theta$ and  $\varphi$. Thus
 	\begin{equation}\label{EspB}
 	E \sb{sp}  (k)=  \frac{k^2}{2\pi^2} F  (k)\,, \  \mbox{for spherical symmetry. }
 	\end{equation}

 	\begin{table*}
 		
 		\begin{tabular}{ ||  c|c|c ||  c |  c|c  |  c||c |c|c|c |c || c|c| c| c||}
 			\hline\hline
 			1    &                     2                     &    3     & 4                                             & 5                       &            6            & 7                      &     8     & 9                    &          10          &          11          & 12                   & 13               &           14           & 15       & 16       \\ \hline
 			$T,$ & $\dfrac{ \white{\big |} \rho\sb n}{\rho}$ & $\alpha$ & $Q, $                                         & $U\sb n,$               &      $U\sb {ns},$       & $ {\cal L},$           & Re$\sb n$ & $k_\times $          &       $k_\nu $       &        $k_* $        & $k_\ell$             & {$n+1$} & $\<m \>_{10}$ & $m\Sb C$ & $m\Sb F$ \\
 			K    &                                           &          & $\dfrac{\rm mW}{  \white{\big |}\rm cm^2}   $ & $\dfrac{\rm cm}{\rm s}$ & $\dfrac{\rm cm}{\rm s}$ & $\dfrac{1}{\rm cm^2 }$ &           & $\dfrac{1}{\rm cm }$ & $\dfrac{1}{\rm cm }$ & $\dfrac{1}{\rm cm }$ & $\dfrac{1}{\rm cm }$ &                  &                        &          &          \\ \hline\hline
 			     &                                           &          & 150                                           & ~1.87~                  &         ~2.32~          & 86300                  &  ~37.9~   & ~37.2~               &         149          &         294          & ~1846~               & 1.89             &          2.00          & ~1.7~    & ~3.0~    \\
 			1.65 &                   0.11                    &   0.11   & 200                                           & 2.23                    &          2.76           & 16200                  &   46.2    & 58.8                 &         224          &         402          & 2529                 & 2.14             &          2.10          & 1.8      & 3.0      \\
 			~    &                                           &          & 300                                           & 3.27                    &          4.04           & 38200                  &   84.9    & 94.5                 &         354          &         618          & 3883                 & 2.18             &          2.20          & 1.9      & 2.9      \\ \hline
 			     &                                           &          & 200                                           & 1.18                    &          1.85           & 81100                  &   53.2    & 43.8                 &         249          &         322          & 1788                 & 1.88             &          1.88          & 1.7      & 3.0      \\
 			1.85 &                   0.19                    &   0.18   & 300                                           & 1.78                    &          2.80           & 19800                  &   94.2    & 70.1                 &         539          &         502          & 2796                 & 2.23             &          1.95          & 1.8      & 2.8      \\
 			~    &                                           &          & 497                                           & 3.03                    &          4.76           & 58500                  &    165    & 123                  &         755          &         863          & 4806                 & 2.35             &          2.20          & 1.9      & 2.8      \\ \hline
 			     &                                           &          & 233                                           & 0.86                    &          1.92           & 14100                  &   84.7    & 73.3                 &         455          &         418          & 2359                 & 2.3              &          2.20          & 1.7      & 3.0      \\
 			2.00 &                   0.55                    &   0.28   & 386                                           & 1.34                    &          3.00           & 47300                  &    131    & 158                  &         690          &         765          & 4321                 & 2.31             &          2.30          & 1.9      & 2.8      \\
 			~    &                                           &          & 586                                           & 2.09                    &          4.67           & 112000                 &    223    & 240                  &         1102         &         1178         & 6650                 & 2.36             &          2.30          & 2.2      & 3.0      \\ \hline
 			~    &                                           &          & 200                                           & 0.57                    &          2.20           & 37300                  &    107    & 170                  &         588          &         612          & 3837                 &  2.30  &          2.25          & 1.7      & 2.9      \\
 			2.10 &                   0.74                    &   0.48   & 300                                           & 0.88                    &          3.40           & 89800                  &    159    & 264                  &         958          &         951          & 5951                 &  2.30  &          2.30          & 2.1      & 3.0      \\
 			     &                                           &          & 350                                           & 0.99                    &          3.83           & 114000                 &    211    & 298                  &         1142         &         1071         & 6705                 &  2.30  &          2.35          & 2.2      & 3.0      \\ \hline\hline
 		\end{tabular}		
 		\caption{\label{t:1}   Columns \#\# 1-3 -- The temperature and temperature-dependent material parameters of \He4 ;\ Columns \#\# 4--7: the experimental  parameters of the    flow.  Column \# 8: the estimates of the normal-fluid Reynolds number, \Eq{kDb}; Columns \# 7--12: the estimates of  the characteristic wavenumbers:  $k_\times$,  $k_\nu$, $k_*$ and $k_\ell$, \Sec{ss:est}. Column  \#13 -- the estimates of the scaling exponents of the energy spectra via apparent scaling exponents of $S_2(Y)$.  Column  \#14 -- the mean-over-decade scaling exponents of the energy spectra $\<m \>_{10}$.	Columns   \# 15 and \# 16: the apparent scaling exponents of the energy spectra in the cascade dominated subinterval, $m\Sb C$ and in the mutual-friction dominated  subinterval, $m\Sb F$.
 		}
 	\end{table*}
 	
  Some   information about possible anisotropy of the $2\sp{nd}$-order statistics of turbulence  can be obtained by comparison of the 1D (spherically averaged) functions  $  E\sb{sp} (k)$, \Eq{EspC} with the 1D  functions $ E_{xy}(k_z)$, $ E_{zx}(k_y)$,  and  $ E_{yz}(k_x)$, averaged over $xy$, $zx$, and $yz$   planes. These functions    depend  only on the   projections of  $\B k$ orthogonal to the corresponding planes. Understanding $F (\B k)$ in the Cartesian coordinates as  $F (k_x,k_y,k_z)$,   we define
 	\begin{eqnarray}  \nn
 	E_{xy}(k_z)&=&\int  F  (k_x, k_y, k_z) \frac{dk_x \, dk_y}{4\pi^2}\,,  \\ \label{EspC}
 		E_{zx}(k_y)&=&\int  F  (k_x, k_y, k_z) \frac{dk_x \, dk_z}{4\pi^2}\,,  \\ \nn
 	E_{yz}(k_x)&=&\int  F  (k_x, k_y, k_z) \frac{dk_y \, dk_z}{4\pi^2}\ .
 	\end{eqnarray}
  	  \end{subequations}

   The total kinetic energy $E$ of the system can be found by respective integration:
 	\begin{equation}  \label{Esp1B}
 	E =  \int   E\sb{sp} (k ) \frac{d k  }{2\pi}=  \int E_{xy}(k_z) \frac{d k_z }{2\pi}=\dots \ .
 	\end{equation}
 	
 In the case of spherical symmetry, all four  1D functions  \Eq{EspB} and   \Eq{EspC}  are proportional to each other (i.e., differ only by numerical prefactors). If the angular distribution of energy is not symmetric, the  behavior of different energy spectra will differ.

 \subsection{\label{ss:new} A new way of statistical analysis of the visualization data}
 As we mentioned, the visualization technique allows one to measure the streamwise velocity across a channel as a function of a wall-normal coordinate $y$ for  fixed values of the  time $t_0$ and the spanwise and streamwise   positions     $z_0$ and  $x_0$:
 \begin{equation}\label{vxy}
 u(y)\= u_x(x_0,y,z_0,t_0)\,.
 \end{equation}
  For simplicity,  we choose $t_0=0$  and $z_0=x_0=0$, i.e., $u_x( 0,y, 0, 0)$.

  So far, the way to statistically analyze the data, i.e., Eq.~\,\eqref{vxy}, was to find the velocity differences $\Delta u_x(Y,y,t)$ defined in \Eq{Snb}
  and calculate the structure functions $S_n (Y)$ using \Eqs{Sna} and \eqref{Snc}.
The theoretical analysis of homogeneous turbulence is traditionally done in the Fourier space, where different Fourier components are statistically independent:  $\< v^\alpha (  \B k)   v^\beta (\B k') \>=0$, if $\B k\ne \B k'$.   We will demonstrate in this paper  that similar approach (in the $ k_y$-space) to the data analysis of the visualization data allows one to get additional information on the statistics of counterflow turbulence that is hidden in the approach based on $S_2(\delta)$.

   To this end, we define a 1D-Fourier transform, $ v (k_y)$ similar to its 3D-version Eq.~\,\eqref{def-Fb}:
 \begin{subequations}\label{1DF}
  \begin{equation}\label{1DFa}
 v (k_y)= \int _{-D/2}^{D/2} u  (y ) \exp (i k_y   y) \,  d y \ .
 \end{equation}
 Here $y=0$ is the position of the centerline and $D$ is the channel width.
   Similarly to  \Eq{def-Fa}, we define next the 1D energy spectrum
\begin{equation}\label{1DFb}
 2\pi\,  \delta (k_y+k_y')  E(k_y)=\< v(k_y) v(k_y') \>\,,
 \end{equation}
 which is nothing but $E_{xz}(k_y)$ defined by \Eq{EspC}. To see this,  notice that integration over $d k_x/(2\pi)$ and $d k_z/(2\pi)$ in \Eq{EspC}, according to \Eq{def-Fc}, results in $\B u_i(x=0,y,z=0)$.

 Our expectation is that  $v (k_y)$ (and respectively $E(k_y)$) better separates turbulent fluctuations with different scales than  $\Delta u(Y,y)$ (and respectively $S_2(Y)$). To see this,  one may consider relation between $E(k_y)$ and $ S_2 (Y)$. Using inverse Fourier transform of \Eq{1DFa}:
 \begin{equation}\label{1DFc}
 u (y)= \int  v  (k_y ) \exp (-i k_y   y) \,  d y \,,
 \end{equation}
 in the definition of  $S_2(Y)$  (i.e., Eq.~\eqref{Sna}) and applying \Eq{1DFb}, one gets:
 \begin{equation}\label{1DFd}
  S_2 (Y)=  \frac 2\pi   \int \sin^2 \Big (\frac{k_y Y}2\Big ) E_{xz} (k_y)\,  d k_y \ .
 \end{equation}
 \end{subequations}
 If this integral converges, it is dominated by the range $k\sim 1/Y$.   Therefore, for the infinite extend of the inertial interval   $S_2 (Y)\sim  F_2(1/Y) Y ^{-1}$. For example, if $F(k)\propto k^{-m}$, then
 $S_2(Y)\propto Y ^n$ with $n=m-1$.  It is important to note that integral \Eq{1DFd} has  also contribution from a wide range of $k$ around $1/Y$. Therefore,  in a realistic situation with a finite extend of available $k$-space,  the relation between $F_n (k_y)$ and $S_n (Y)$ is not so simple.

In any case, one expects, as we will demonstrate in this paper, that direct measurement of the \textit{integrand }$E_{xz}(k_y)$ gives more detailed information about the statistics of counterflow turbulence than the measurements of the \textit{integral} (Eq.~\,\eqref{1DFd}) for  $S_2(Y)$.

  \subsection{\label{ss:theory} Overview of the theory of counterflow turbulence}
  Analytical theory of counterflow superfluid turbulence, developed in  \Ref{LP-2018}, describes the
  energy spectra of the normal-fluid and superfluid components of superfluid $^4$He at scales $r$ exceeding intervortex distance $\ell=1/\sqrt{\C L}$,  where $\C L$ is the vortex density, i.e., total length of vortex lines per unit volume.   The theory is based   on  the gradually-damped version\cite{Tlambda} of the
  coarse-grained  Hall-Vinen-Bekarevich-Khalatnikov(HVBK)
  equations, generalized in \Ref{decoupling} for the counterflow turbulence. These   equations  have  a
  form of two Navier-Stokes equations  for the turbulent velocity fluctuations $\B
  u\sb n(\B r,t)$ and $\B u\sb s(\B r,t)$, coupled by a simplified version of the mutual friction force\cite{LNV}
  \begin{equation}\label{mf}
     \B f\sb{ns}\simeq
 \Omega\sb s\,  \big [\B u\sb n(\B r,t)- \B u\sb s(\B r,t)\big ]\,, \quad \Omega\sb s =\alpha (T)\kappa \C L\, .
  \end{equation}
  Here $\alpha(T)$ is the temperature dependent parameter of the mutual friction, listed in Table\,\ref{t:1}, column \#\,3.

  These equations served as a starting point for derivation of  the stationary balance equations for the 1D energy spectra
   $E\sb n(k)$ and $E\sb s (k)$ of the normal and superfluid components
   \begin{equation}\label{balance}  \frac{d \ve_j(k)}{d k} = \Omega_j \big
   [E\sb{ns}(k) - E _j (k) \big ] - 2\, \nu_j k^2 E_j(k)\,,
   \end{equation}
   using simplifying assumption of the  spatial homogeneity and isotropy of the counterflow turbulence statistics.
   Here $\ve _j (k)$ is the energy flux over scales  $1/k$  in   the normal-fluid ($j=$n) and superfluid  ($j=$s) velocity components,  $\Omega\sb n = \Omega\sb s \rho\sb s/  \rho \sb n$, $\nu\sb n$ is the normal-fluid kinematic viscosity (normalized by the $\rho\sb n$), and $\nu\sb s$ is   the Vinen's effective superfluid viscosity\,\cite{Vinen}. The viscous-like energy sink term  was added to HVBK equations in \Ref{Tlambda} to account for the energy dissipation at the intervortex scale $\ell$ due to vortex reconnections, the energy transfer to  Kelvin waves, and similar effects.
   In \Eq{balance}, $E\sb n(k)$ and  $E\sb s(k)$ are the 1D spherically averaged energy spectra  [cf. \Eq{EspA}] of the normal-fluid and superfluid subsystems and the cross-correlation function $E\sb{ns}(k)$ is related similarly to $F\sb{ns}(\B k)$.

 Eqs.\,\eqref{balance} are exact (in the framework of HVBK equations), but not closed. To make them practically useful, the closure approximations for $\ve_j(k)$ and for $E_j(k)$ in the terms of $E_j(k)$ were used.

  The role of long-range (in the $k$-space) energy-transfer terms was analyzed\cite{LP-2018},  based on the integral closure for $\ve(k)$\,\cite{LNR},  and  a new self-consistent closure was suggested:
 \begin{subequations}\label{LP17}
 	\begin{equation}\label{LP17a}
 	\ve(k)= C(k) k ^{5/2} E^{3/2}(k)\,, \  C(k)= \frac{4\, C}{3\,[3-m(k)]}\ .
 	\end{equation}
 	in which $m(k)$  should be understood as a  local scaling exponent of $E(k$):
 	\begin{equation}\label{LP17b}
 	m(k)=\frac{d \ln E(k)}{d \ln (k)} \ ,
 	\end{equation}
 \end{subequations}
and the prefactor $C(k)$ is chosen to reproduce the Kolmogorov constant
   	$C$ for the K41 scaling exponent $m(k)=5/3$.

   To complete the closure of \Eqs{balance}, the closure\cite{decoupling}  for  the cross-correlation function $E\sb{ns}$ was adopted. In a simplified form, suitable for conditions of the visualization experiments\,\cite{WG-2015,WG-2017}, it reads:
   \begin{subequations}\label{Ens1}
   	\begin{eqnarray}\label{Ens1A}
   	&& \hskip -.3  cm E\sb{ns}(k)=D(k)E\sb{ns}^{(0)}(k)\,, \ D( k)= \frac{\arctan
   		[\xi(k)]}{\xi(k)}\,, ~~~~~~~~\\
   	&&  \hskip -.3  cm \xi(k)= \frac{k}{k_\times}\,, \   k_\times= \frac{\Omega\sb{ns}}{U\sb{ns}}\,, \ \Omega\sb{ns}=\Omega\sb n+ \Omega \sb s= \alpha \kappa \C
   	L\frac{\rho}{\rho\sb n}  \,, ~~~~~~~~ \\  \label{Ens1D}
   	&&  \hskip -.3  cm  E\sb{ns}^{(0)}(k)=    \big [\rho\sb n E\sb n(k)+
   		\rho\sb s E\sb s(k)\big ] / \rho\ .
   		\end{eqnarray}\end{subequations}
   	Here  $D(k)$ is the $U\sb {ns}$-dependent decoupling function and  $E\sb n(k)$ and $E\sb s(k)$  are  the $U\sb{ns}$-dependent energy spectra, found
   self-consistently by solving \Eqs{balance} with $E\sb {ns}(k)$  given by \Eqs{Ens1}.

   Further simplification of the balance \Eqs{balance}, \eqref{LP17}, and \eqref{Ens1} for the experimental  conditions, results in decoupled balance equations for the normal-fluid and the superfluid energy spectra:
   	\begin{eqnarray}\nn
   	C(k) \, \frac d {dk} k^{5/2} E_j ^{3/2}(k)&=& E_j (k)\big \{ \Omega_j [ D(k)-1]\\ \label{balance2}
   	\label{fin-bal} && - 2 \nu_j k^2 \big \}  , ~~~~~~
     	\end{eqnarray}
  in which $C(k)$ and $D(k)$ are given by \Eqs{LP17a} and \eqref{Ens1A}.
 The solutions of these equations are compared with the experimental spectra in \Figs{F:3}.\vskip .3cm

  To summarize this overview, note that analytical theory\,\cite{LP-2018}  describes the main features of the large scale normal-fluid energy spectra of counterflow turbulence, observed in the visualization experiments,  although it does not account for the inhomogeneity and anisotropy of the flow.

  \begin{figure*}
  	\begin{tabular}{ccc}
  	(a)\ & (b)\  & (c)\  \\
  	\includegraphics[scale=0.36]{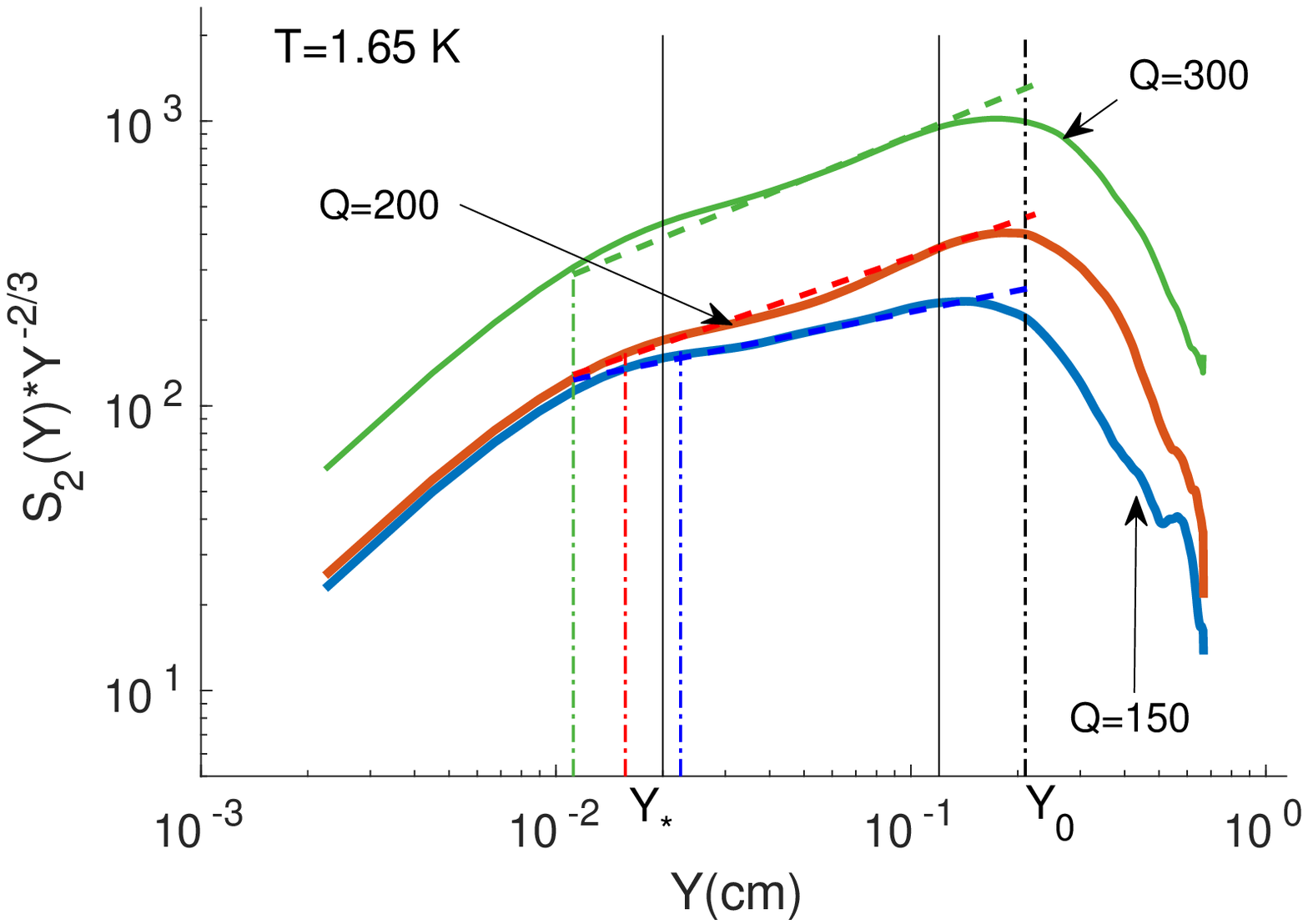}&
  	\includegraphics[scale=0.37]{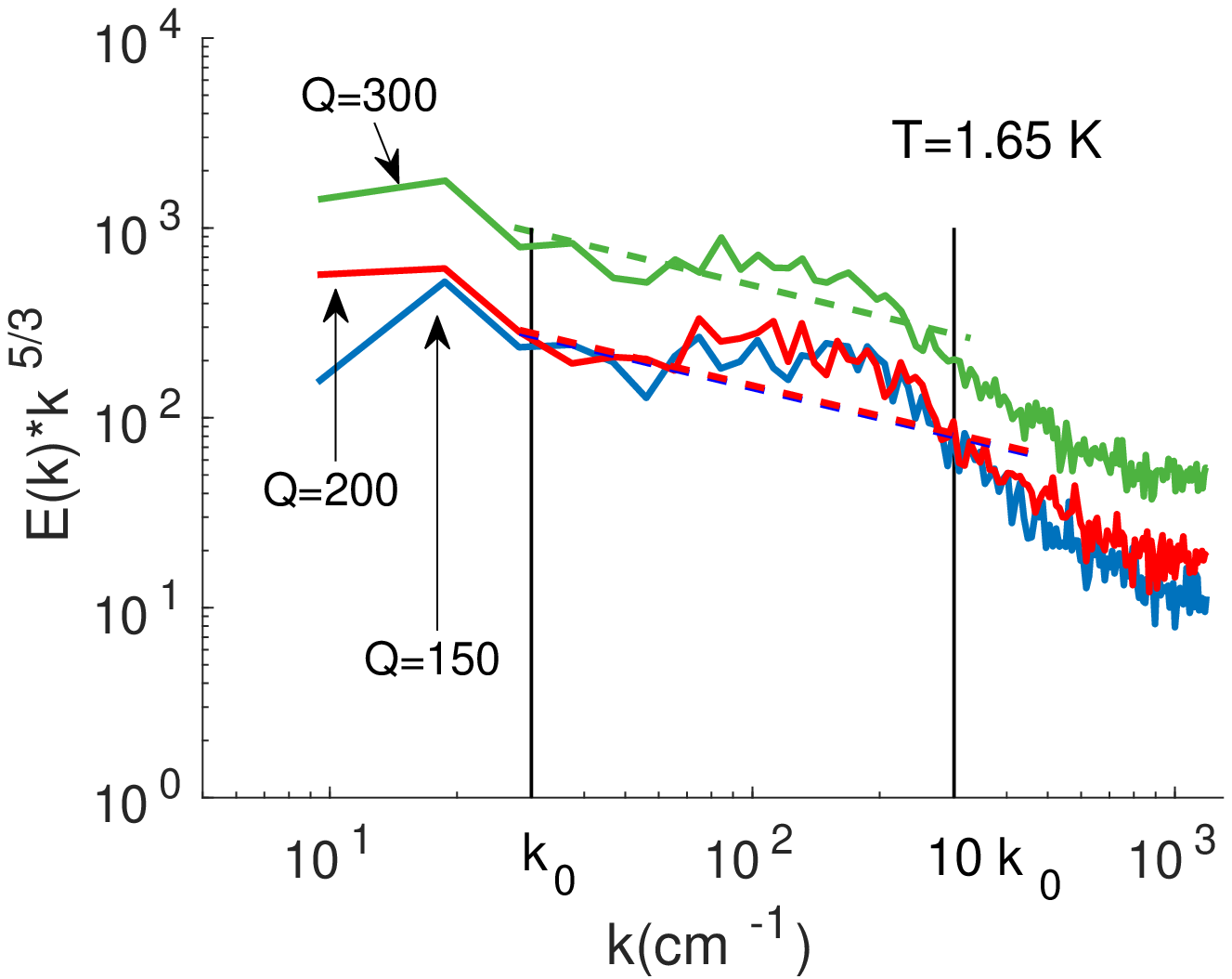} &
  	\includegraphics[scale=0.33 ]{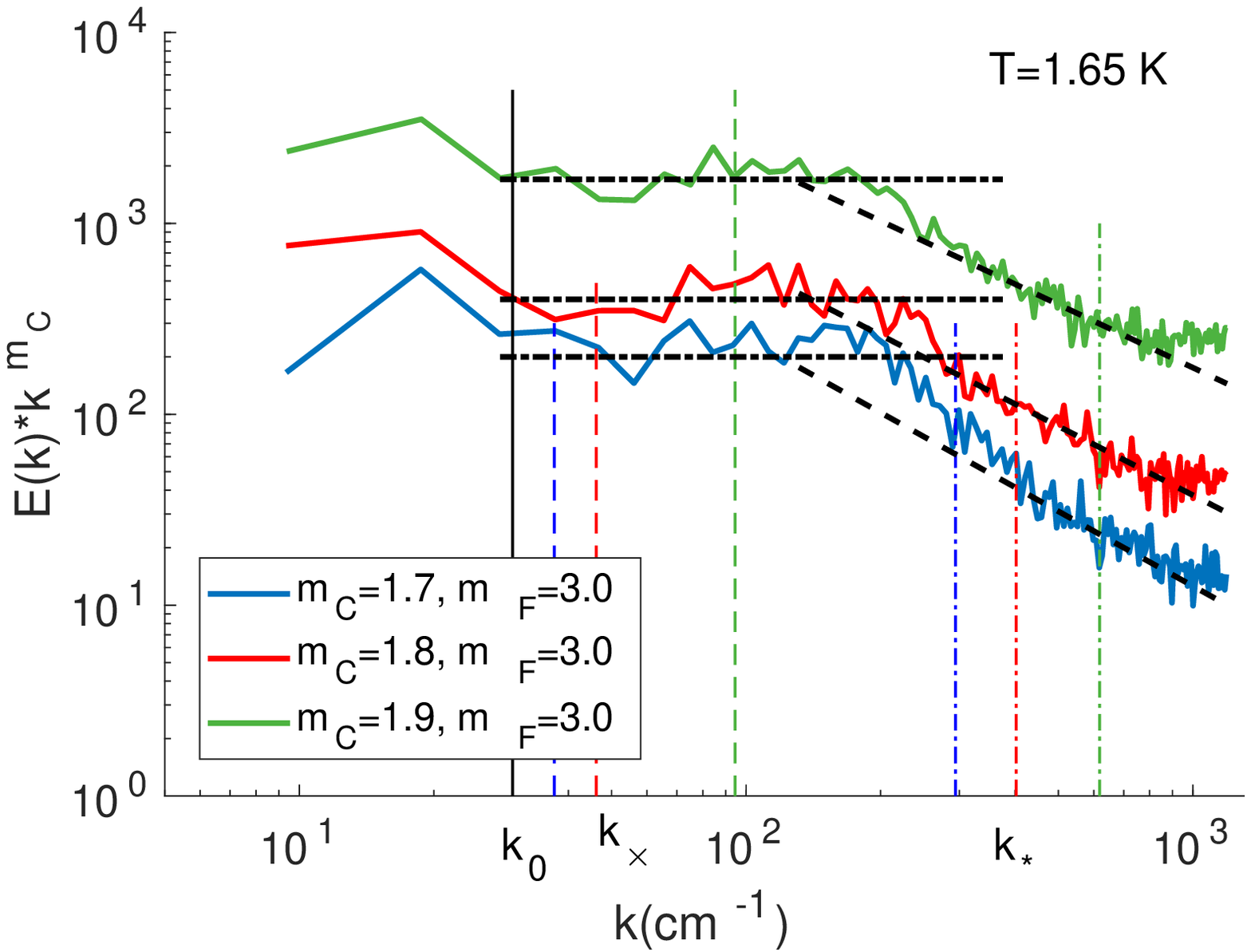}\\
  	(d)\ & (e)\  & (f)\ \\
  	\includegraphics[scale=0.36]{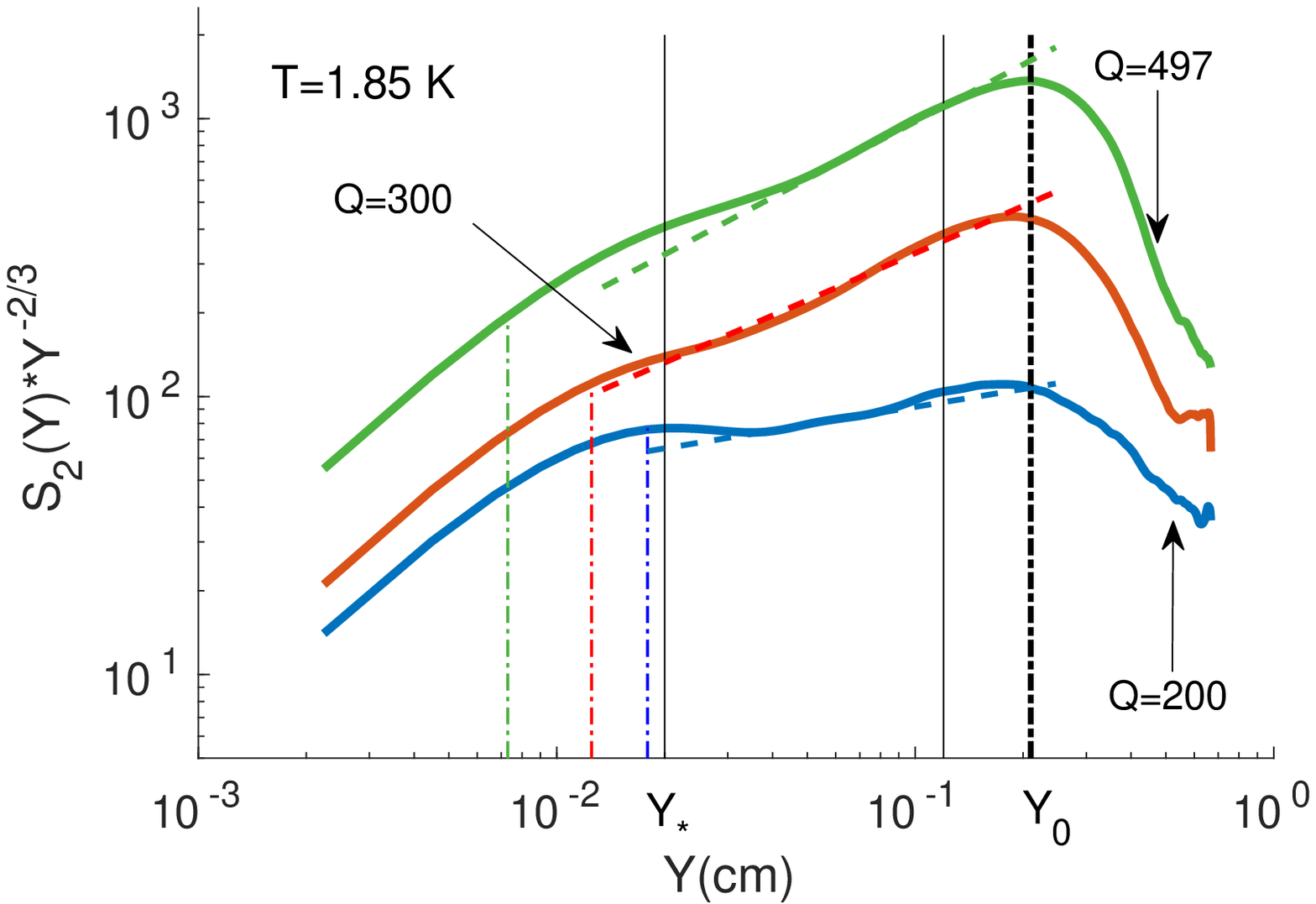}&
  	\includegraphics[scale=0.37]{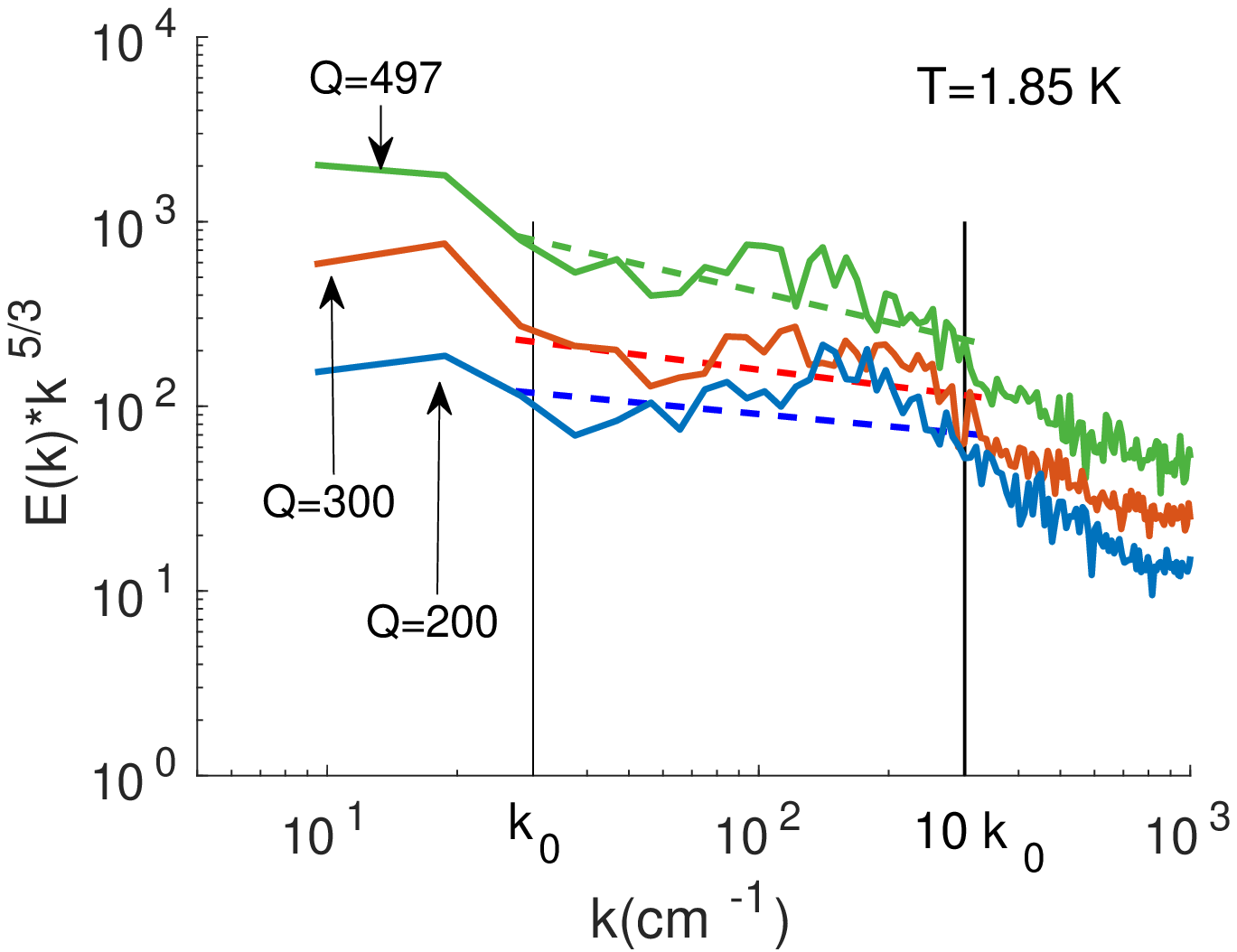} & 	
  	\includegraphics[scale=0.32 ]{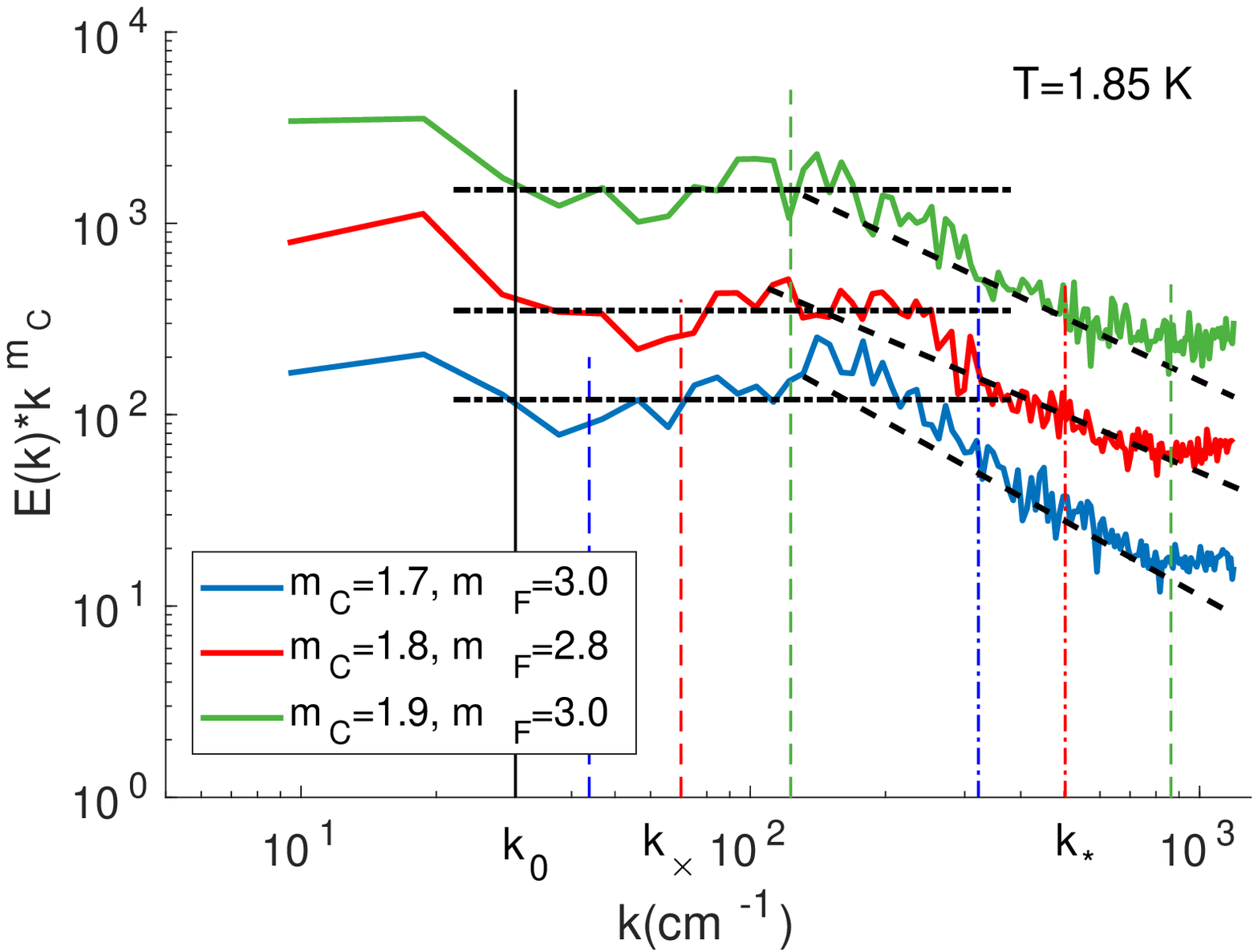}  \\
  	(g) & (h)\ & (i)\  \\
  	\includegraphics[scale=0.36]{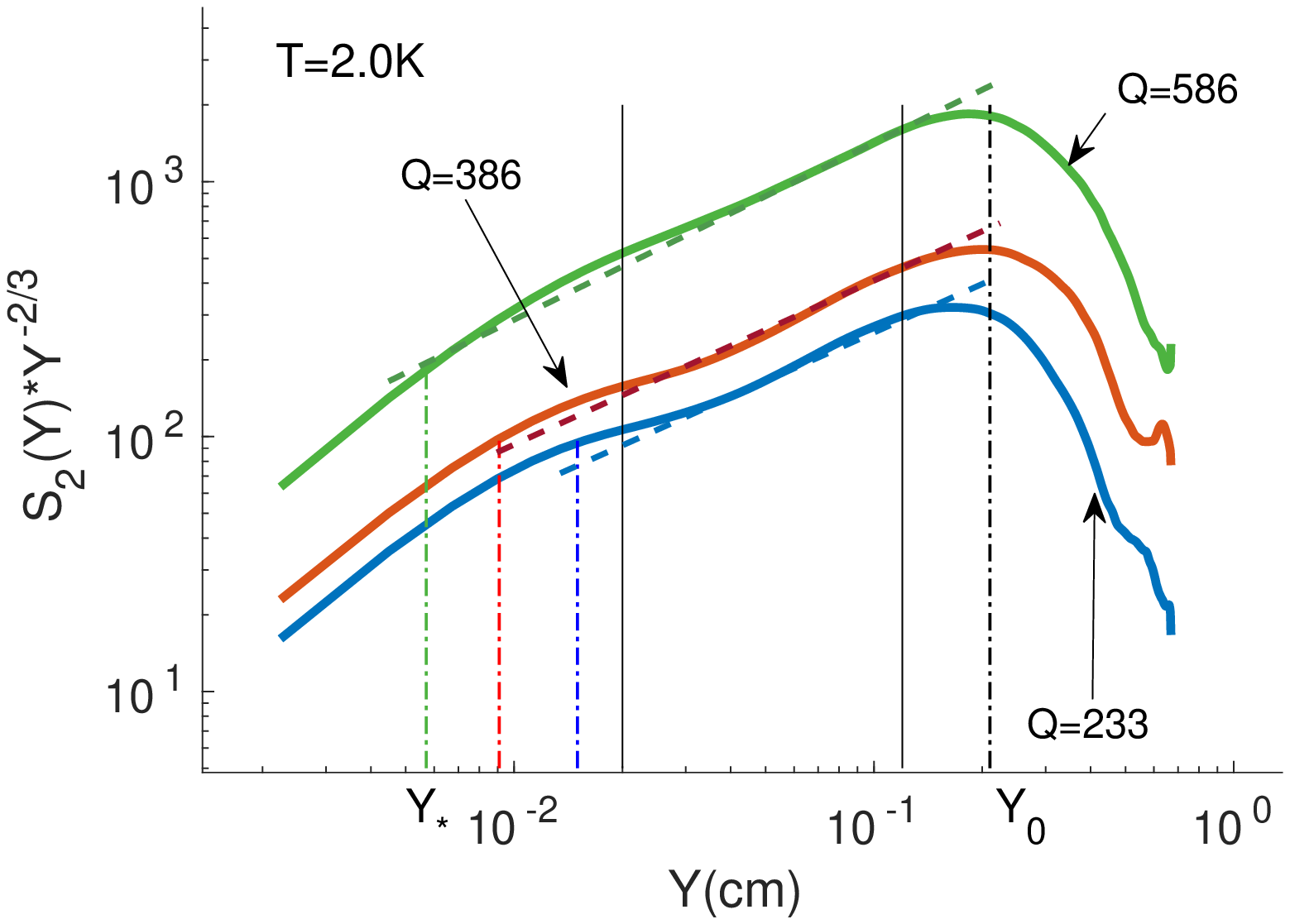}&
  	\includegraphics[scale=0.37]{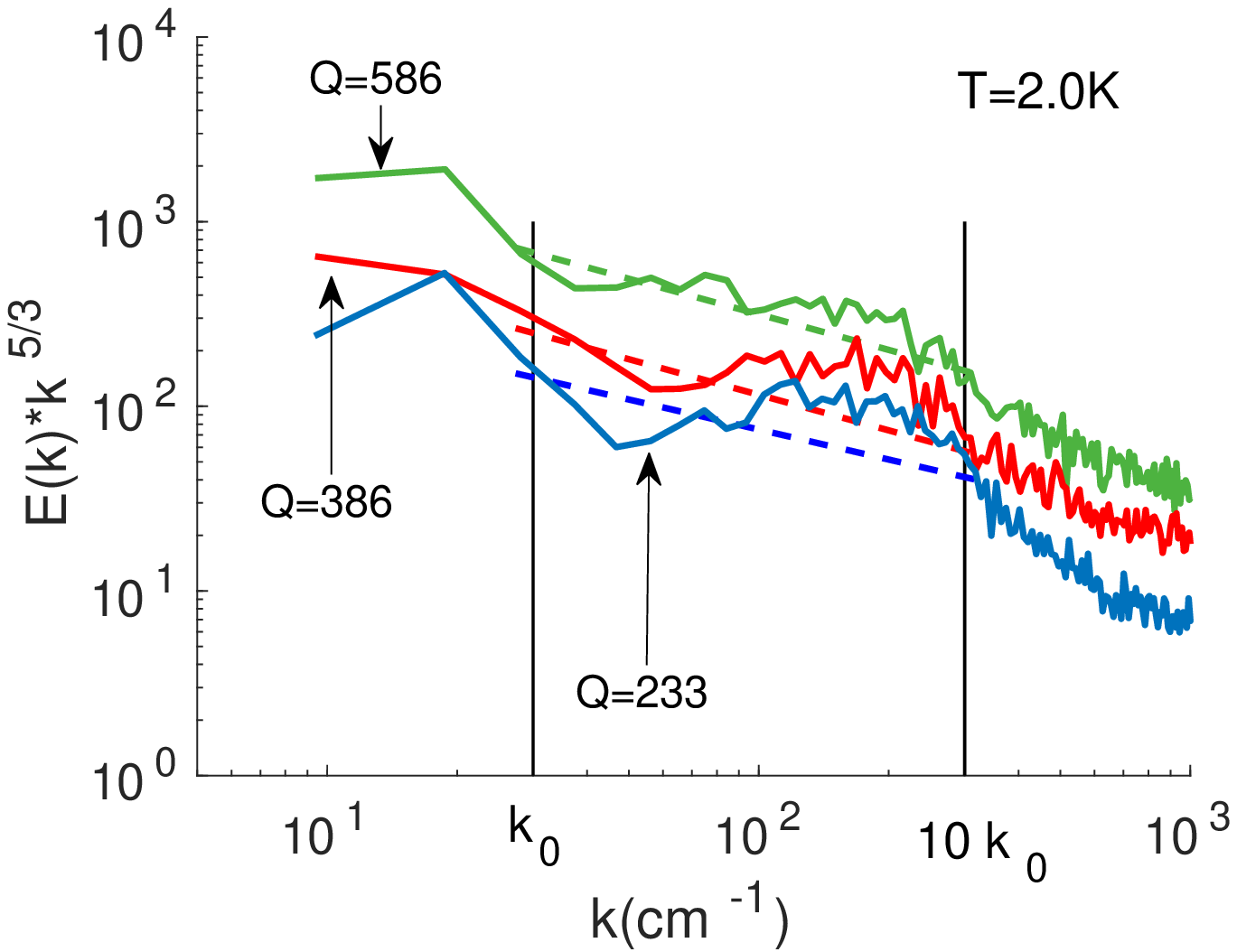} &
  	\includegraphics[scale=0.33 ]{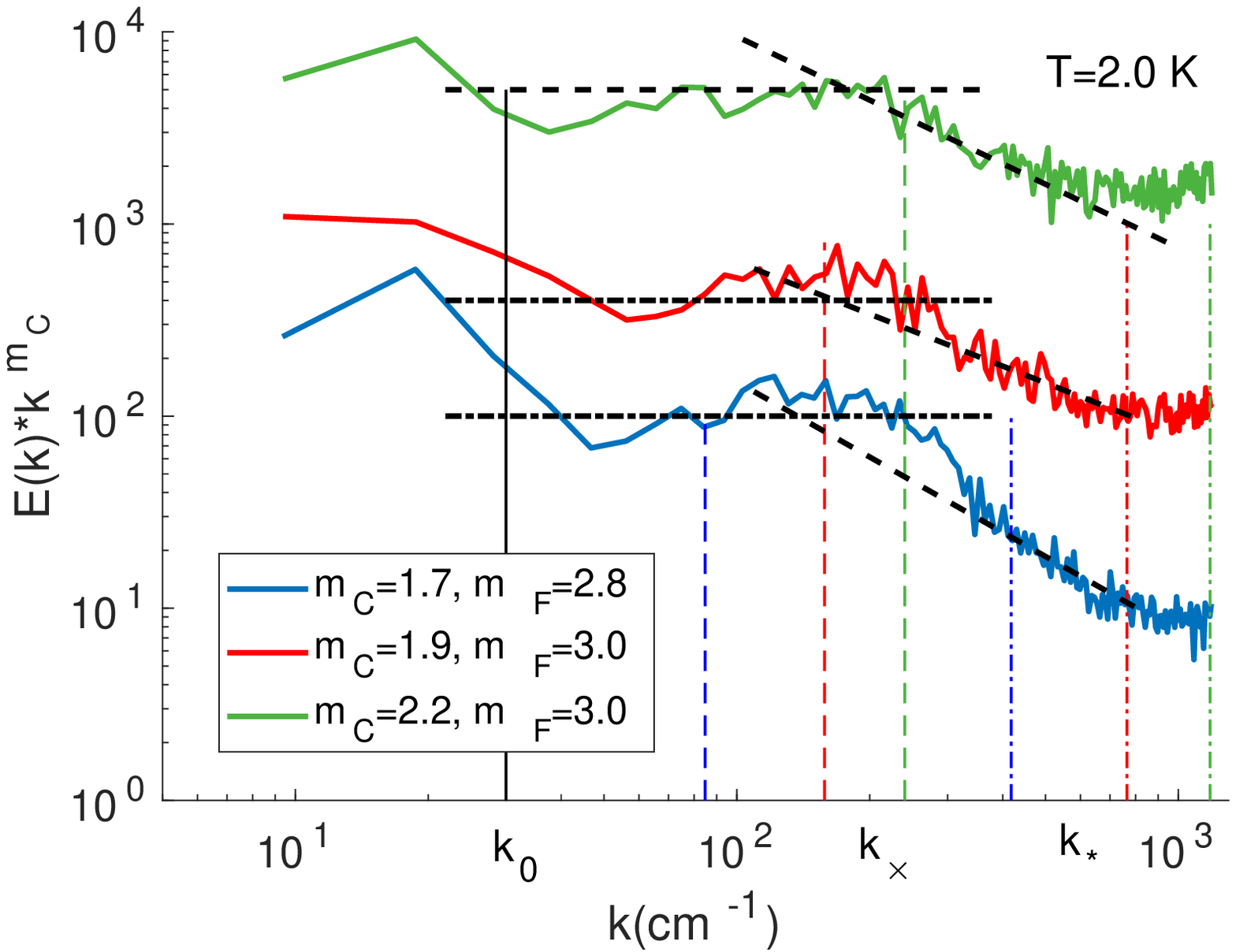}\\
  	(j)\  & (k)\ & (l)\   \\
  	\includegraphics[scale=0.36]{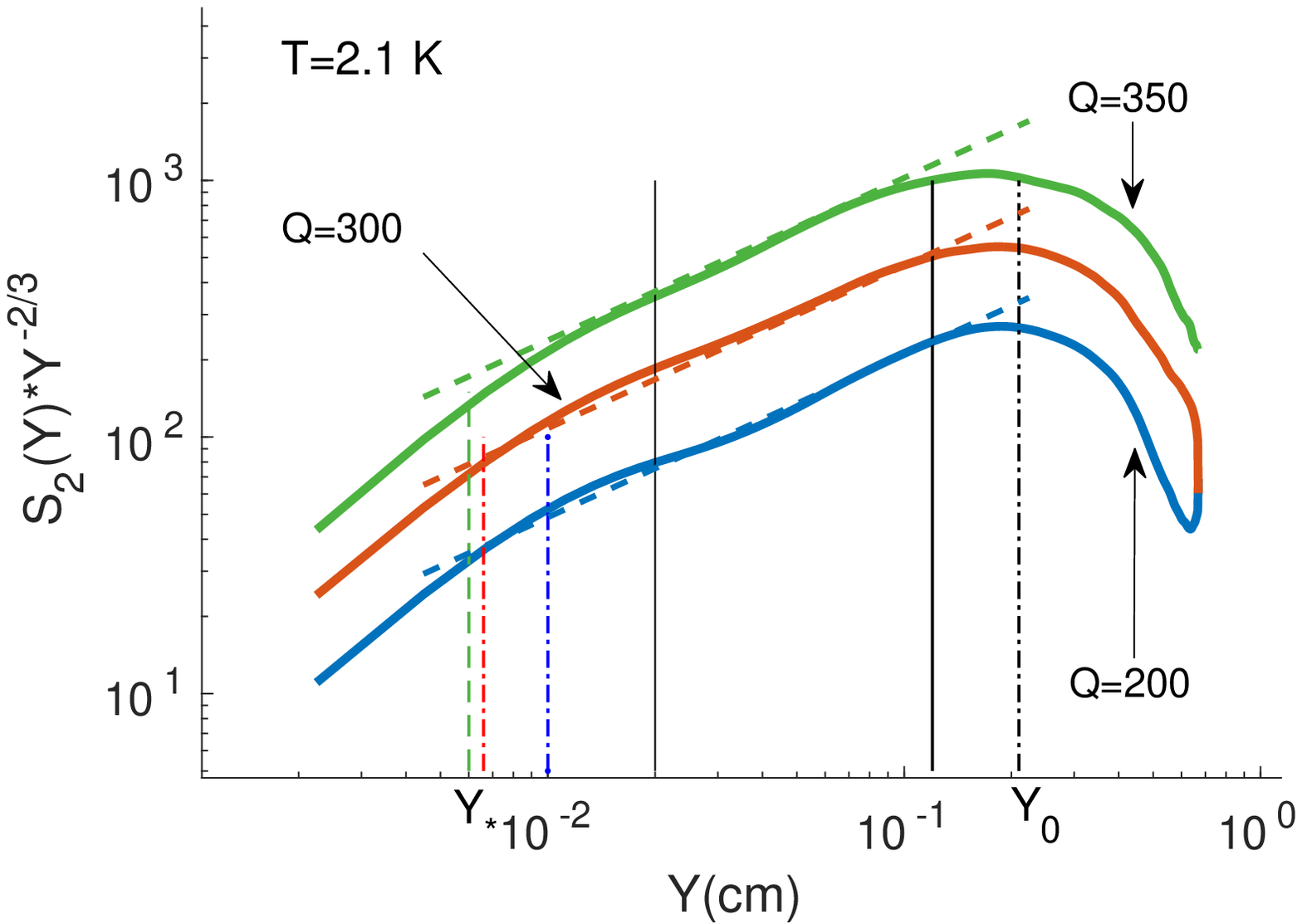}&
  	\includegraphics[scale=0.36]{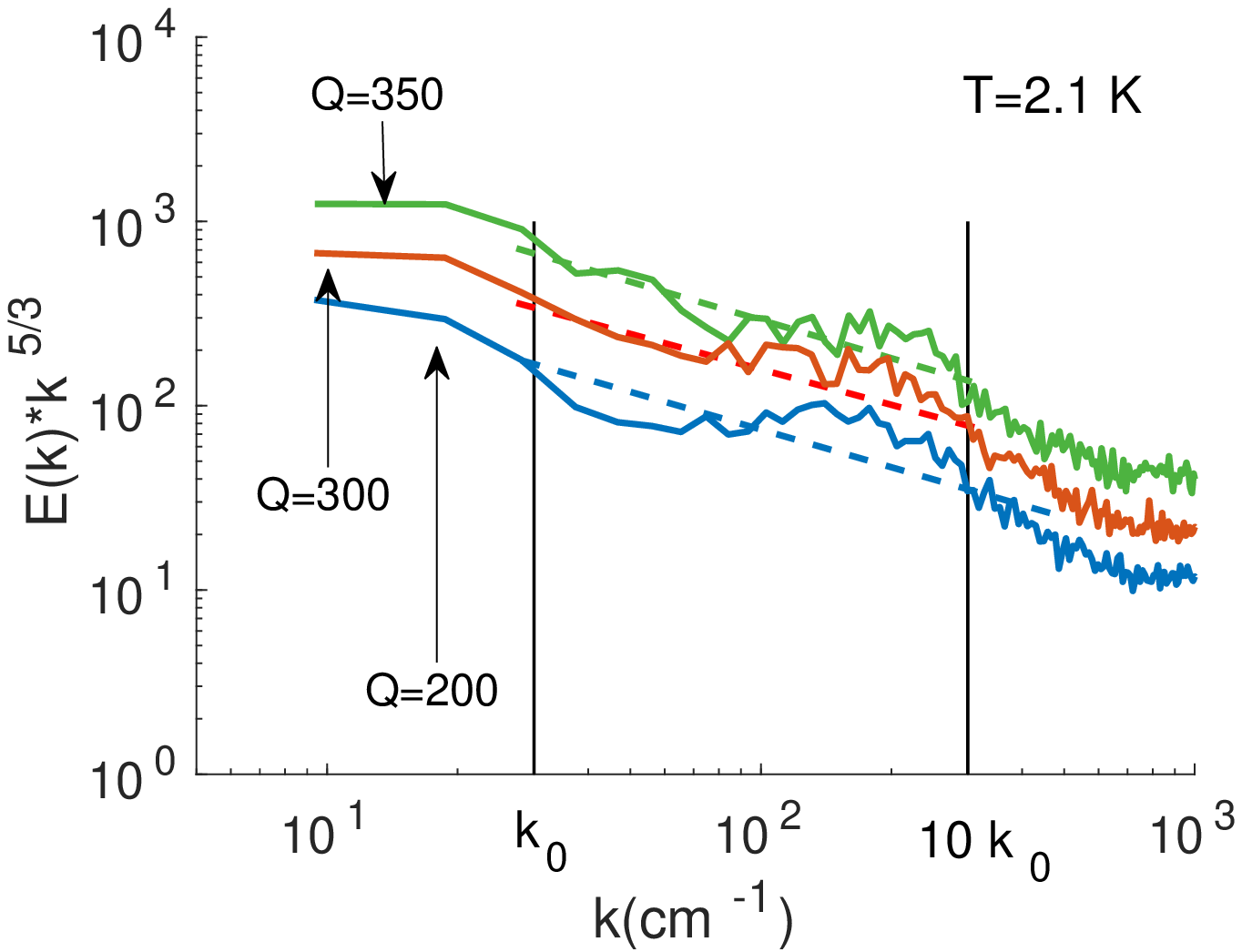} &
  	\includegraphics[scale=0.33 ]{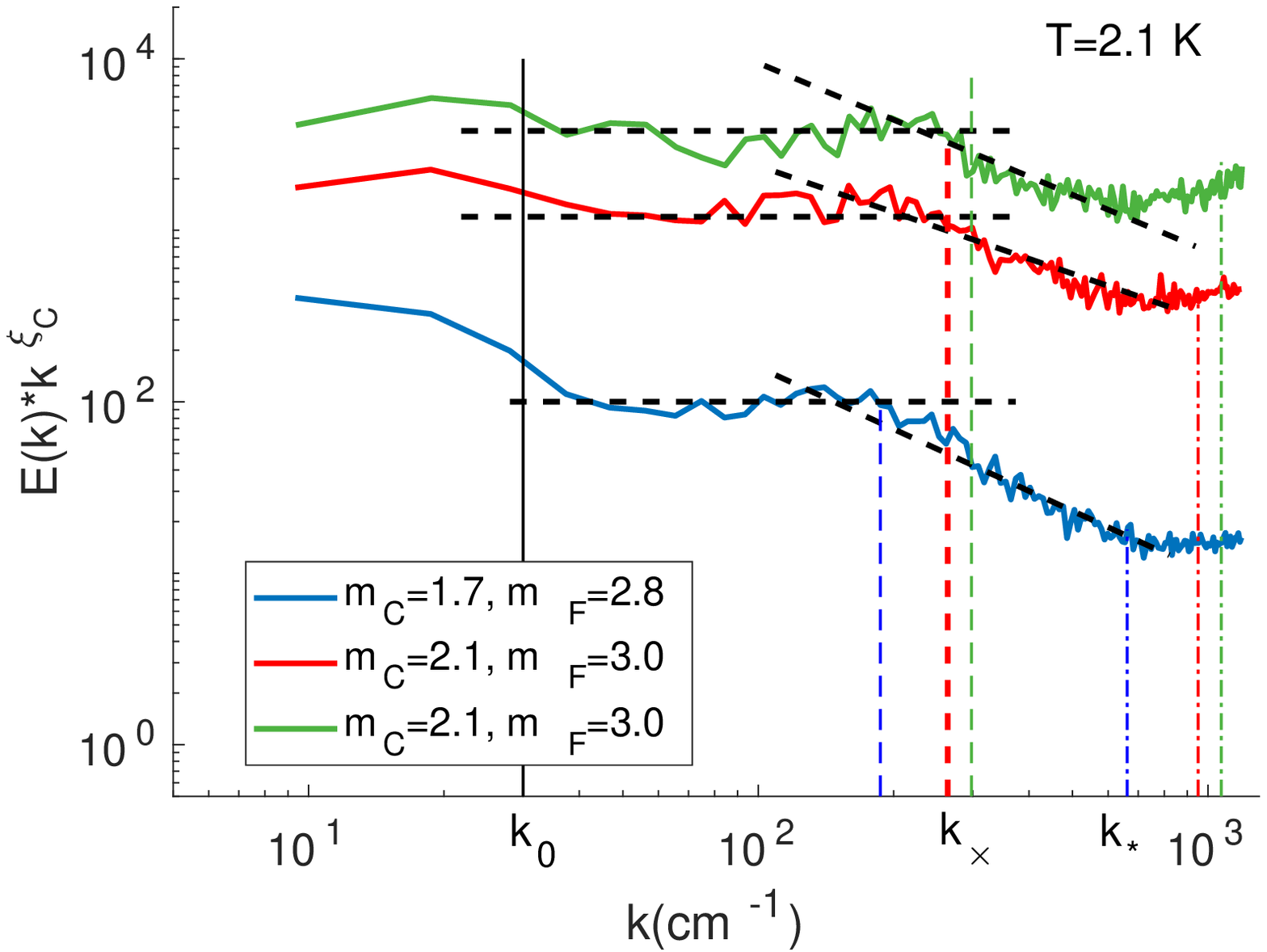} \\
  \end{tabular}
  	\caption{  \label{F:1}  The second-order statistics  for different $T$ and heat fluxes $Q$.  The figures in the rows (top to down) correspond to $T=1.65$~K,  1.85 K, 2.0 K, and 2.1 K, respectively.  Left column: The second-order structure functions, compensated by K41 scaling,  $Y^{-2/3}S_2(Y)$.  The colored thin  straight  lines denote fits with exponents $n$ (cf. Table\ref{t:1}, Column \#13). The fitting range (according to  \Ref{WG-2017}) is denoted by black vertical thin lines. The black vertical dot-dashed line marks the outer scale of turbulence, denoted as $Y_0$.  The colored vertical dot-dashed line (collectively marked as $Y_{*}$) denote the scales, corresponding to the respective  crossover wavenumbers $k_{*}$.  Middle column:  the energy spectra compensated by K41 scaling $k^{5/3}E(k)$. Dashed lines of matching colors denote fits $E(k)\propto k^{-\<m\>_{10}}$ in the wavenumber interval $k \in [k_0$ to $10\, k_0]$, shown by  black vertical lines. Right column: the energy spectra, compensated by an individual scaling, $k^{m\Sb C}E(k)$, found by  fitting  each spectrum in the cascade-dominated range. The compensation  is emphasized by horizontal  dot-dashed lines. The  fits of the mutual-friction dominated range, $E(k)\propto k^{-m\Sb F}$ are shown by black dashed lines. The vertical lines, corresponding to the crossover wavenumbers $k_{\times}$(dashed lines) and $k_*$(dot-dashed lines), are marked in the same color as the corresponding spectra. The legend indicate the scaling exponents $m\Sb C$ for the cascade-dominated range and $m\Sb F$ for the mutual-friction dominated range. The outer scale of turbulence $k_0= 2\pi /Y_0  $ is marked as a black solid thin line.  Different  heat fluxes $Q$ (in mW/cm$^2$) are color coded:  green lines  denote the largest $Q$, red lines -- intermediate $Q$ and blue lines -- the smallest $Q$. The color code  is the same in all panels. }
  \end{figure*}

 \section{\label{s:analysis}Experimental results and their analysis }
  In this section we analyze the experimental data of the normal-fluid turbulent velocity fluctuations $u(y)$, obtained  for $T=1.65\,$K,  $1.85\,$K,  $2.00\,$K, and   $2.10\,$K with three  values of heat fluxes $Q$ at each temperature. Main parameters of these experiments are given in Table\,\ref{t:1}.  This table also collects the values of important   characteristic wavenumbers, estimated below in  \Sec{ss:est}.  The second-order statistics (i.e., the structure functions and the spectra $E_{zx}(k_y)$) are discussed in \Sec{ss:E2} and the higher-order statistics of the structure functions in \Sec{ss:flat}.
  
 \subsection{\label{ss:exp} Experimental techniques}
  	The experimental apparatus is identical with that described in \Refs{WG-2015,WG-2017}. A stainless steel channel of 9.5 mm square cross section with a total length of 300 mm is attached to a pumped helium bath whose temperature can be controlled within 0.1 mK by regulating the vapor pressure. A planar heater (around
  	400 $\Omega$) at the lower end of the channel was used to drive a thermal counterflow. When the heat flux is sufficiently large, both the superfluid and the normal fluid components can become turbulent. To probe the normal fluid turbulence, our recently developed He$_2^*$ molecular tracer-line tracking technique\,\cite{E3} was adopted. A 35-fs pulsed laser with a repetition rate of 5 kHz and a pulse energy of about 60$\,\mu J$ was focused into the channel to produce a thin line of He$_2^*$ molecular tracers. This tracer line can be driven to produce 640 nm fluorescent light by a pulse train from an imaging laser at 905 nm (i.e., 5-6 pulses at a repetition rate of 500 Hz). The fluorescence was captured by an intensified CCD (ICCD) camera, mounted perpendicular to both the flow direction and the laser beam path, to produce the images of the tracer line. In a typical experiment, a straight baseline image is acquired to serve as a reference. Then, the heater is turned on for at least 20\,s so that a fully developed counterflow can establish in the channel. After that, we produce a tracer line and let the tracer line move with the normal fluid by a drift time $\Delta t$ before the drifted line is imaged.
  	
  	In order to extract quantitative velocity field information, the center location of every line segment needs to be accurately determined. In our previous research\,\cite{WG-2015,WG-2017}, a simple Gaussian fit method was adopted. First, the image of a tracer line was cut into many small segments (i.e., typically 40-60 segments). Then, the fluorescence intensity profile of each line segment was fit by a Gaussian function such that both the center location and the width of the line segment can be determined. The streamwise velocity of the normal fluid   at position y can be calculated as the displacement of the line segment divided by the drift time $\Delta t$. This method works well only when the tracer-line image has good quality and high signal-to-noise ratio. However, as the normal fluid velocity increases, some line segments can distort and smear, which can result in significant uncertainty in locating the center of these segments using the Gaussian fit method. In this research, we utilized a more sophisticated approach, which is based on the algorithm proposed by Pulkkinen \textit{et al}. for finding curvilinear structures in noisy data\,\cite{E4}. There are two steps involved in the image analysis. First, a tracer-line image is noise-filtered based on the intensity of bright pixels using numerically inexpensive nearest-neighbors searches. The basic idea is to remove those bright pixels that are surrounded all by dark pixels and hence are more likely created due to instrument or environmental noises\,\cite{E5}. Subsequently, the algorithm of  \Ref{E4} is applied to determine the ridge line of the entire fluorescence intensity profile. The displacement of the ridge line then allows us to calculate the streamwise normal-fluid velocity regardless of the bad quality of some local line segments.

   Based on the obtained streamwise normal-fluid velocity  $u_x(y)$, we can evaluate the velocity difference  $\Delta u_x(Y,y)=u_x(y+Y)-u_x(y)$,    between two line segments that are separated by a distance $Y$. Then, the transverse structure functions of the normal fluid turbulence can be easily computed as  $S_n(Y)=\<|\Delta u_y(y,Y)|^n\>$ , where the angle brackets $\< \dots \>$ denote an ensemble average over all the images obtained under the same experimental conditions (typically 30-100 images). The calculated structure function profiles are found to be insensitive to the reference location $y$. The 1D-energy spectra, averaged over $x$-$z$ plane and parallel to the channel wall can also be determined. In \Fig{F:1}, we show the obtained structure function and energy spectra curves at various temperatures and heat fluxes. It should be noted that the results for separation distance $Y$ smaller than the thickness of the tracer line (i.e., about 100-200 $\mu$m) can have large uncertainties.

  \subsection{\label{ss:est}Estimates of the cross-over wavenumbers}

 \subsubsection{\label{sss:kX} Decoupling  wavenumber $k_\times$}
 According to \Ref{decoupling}, the decoupling wavenumber $k_\times$, for which the decoupling function $D(k)=\frac12$, is  estimated as
 \begin{equation}\label{kD}
 k_\times \simeq 2 \, \Omega\sb{ns}/U\sb{ns}\simeq \kappa \C L/ U\sb{ns}  \ .
 \end{equation}
 For typical values $\C L\simeq 10^5\,$cm$^{-2}$, $U\sb{ns}\simeq 1\,$cm/s and with $\kappa\simeq 10^{-3}\,$cm$^2$/s, this gives $ k_\times\simeq 100\,$cm$^{-1}$. The particular values of $k_\times$ for each of the 12 experimental sets are presented in  Table\,\ref{t:1}, column  \#9 .

 \subsubsection{\label{knu} Viscous wavenumber $k_\nu$}
 The viscous  wavenumber $k_\nu$, for which the viscous damping becomes comparable  with the energy transfer over scales, can be estimated by comparison of the viscous damping frequency $\nu k^2$ with the eddy-turnover frequency $\sqrt{k^3 E(k)}$:
 \begin{subequations}\label{kD}
 	\begin{equation}\label{kDa}
 	k_\nu \simeq E(k_\nu)/\nu^2\ .
 	\end{equation}
 	Using  K41-estimate for the energy spectrum $E\Sb{K41}(k)\simeq u\Sb{T}^2 k_0^{2/3}k^{-5/3}$ we get
 	\begin{equation}\label{kDb}
 	\frac{k_\nu}{k_0} \simeq \Big [\frac{E(k_\nu)}{E\Sb{K41}(k_\nu)}\Big ]^{3/8}\mbox{Re}^{3/4}\,,  \quad \mbox{Re}=\frac{u\Sb T}{k_0 \nu}\ .
 	\end{equation}
 \end{subequations}
Here $k_0\simeq 30\,$cm$^{-1}$, estimated in \Ref{WG-2017} from the behavior of $S_2(Y)$. Our estimates  of $k_\nu$ are given in  Table\,\ref{t:1}, column  \#10.

  Eq.\,\eqref{kDb} is a generalization of the well known K41 relationship $k_\nu\simeq k_0 \mbox{Re}^{3/4}$ for the spectra that differ significantly  from the K41 scaling $E\Sb{K41}(k)\propto k^{-5/3}$.

 \subsubsection{\label{sss:kF}Mutual friction -- viscous crossover wavenumber $k_*$}
 We know that the characteristic frequency, responsible for the rate of  energy dissipation by mutual friction  in the normal fluid component  is $\alpha \, \dfrac{\rho \sb s}{ \rho \sb n} \kappa \C L$, while the corresponding frequency for the viscous dissipation is $\nu\sb n k^2$. Comparing these two frequencies, one find a crossover wave number $k_*$ for which the efficiency of these two dissipation mechanisms is equal:
 \begin{equation}\label{kstar}
 k_* = \sqrt{\alpha \, \dfrac{\rho \sb s}{ \rho \sb n} \frac{\kappa}{\nu\sb n } \C L}\ .
 \end{equation}
 Substituting the particular temperature dependent values of $\alpha$, $\rho\sb n$ and $\nu\sb n$, we get the values shown   in   Table\,\ref{t:1}, rows  \#11. As we see, for $T=1.65\,$K and  $T=2.10\,$K  $k_*\approx 1.0 \, \sqrt {\C L}$, while   for $T=1.85\,$K and  $T=2.00\,$K $k_*\approx 1.1 \sqrt {\C L}$. This is smaller than the wavenumber
 \begin{equation}\label{kell}
 k_\ell \approx 2 \pi \sqrt {\C L}\,,
 \end{equation}
 (cf. Table \,\ref{t:1}, column \#12), that separates the  quasi-classical and  ultra-quantum regimes of superfluid turbulence.

  \subsection{\label{ss:E2}Second-order statistics of counterflow turbulence }
  Fig.\,\ref{F:1} summarizes the second order statistics of the velocity fluctuations for different temperatures and flow parameters. Both the structure functions (left column) and the 1D energy spectra,  are compensated by the  K41 scaling:    $Y ^{2/3}S_2(Y)$,  $k_y^{5/3}E_{xz}(k_y)$.  In the right column, we plot the energy spectra compensated in the cascade-dominated range, see below.

    \subsubsection{\label{sss:S2} Second-order structure functions}
    The  structure functions in the counterflow share similarity with the velocity structure functions in the classical hydrodynamic turbulence.  The expected inertial interval of scales $\delta\sb{min}\approx 0.02\,$cm to $\delta\sb{max}\simeq 0.2\,$cm is marked by black thin vertical lines and correspond to that in \Ref{WG-2017}. Clearly, the structure functions are steeper than their classical counterparts.
  The apparent scaling behavior in this interval of scales may be characterized by exponents $n$. These exponents were found in \Ref{WG-2017} and are reproduced in Table\,\ref{t:1}, column \#13.   The values of $n$ for $T=2.1$\,K are slightly larger than in \Ref{WG-2017},  likely due to the improved image analysis and the fitting procedure.   Note that the values of $n$ vary widely, depending on the flow parameters: the temperature and the heat flux.

    The dot-dashed vertical lines, colored as the structure functions and collectively  marked  $Y_*$ , denote the scale that delineate the ranges of dominance of two dissipative mechanisms: the mutual friction  (for $Y>Y_*$) and the viscous dissipation (for $Y<Y_*$).

    We should also  note  that simple analysis of \Eq{1DFd} shows that the small scale behavior  $S_2(Y)\propto Y ^2$  appears if the energy spectrum $E(k)$ decays as $k^{-3}$ or faster (including the exponential decay). Therefore, the apparent $Y^2$-scaling cannot be  uniquely connected with the viscous dissipation of turbulent kinetic energy, as in the classical turbulence.  Moreover, the asymptotic slope  $S_2(Y)\propto Y ^2$  at  $Y \lesssim Y_*$  is not reached in our experiments due the limited spatial resolution (cf. \Fig{F:4}a).

    \begin{figure*}
    		\begin{tabular}{cc}
    			(a)& (b)\\
    	   	\includegraphics[scale=0.8]{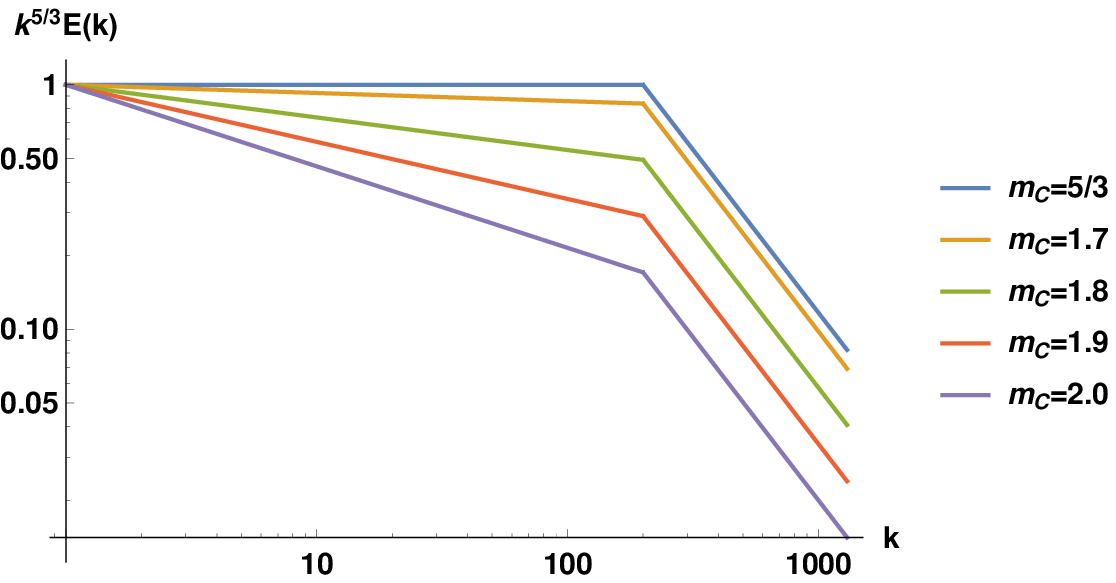} &
    	\includegraphics[scale=0.6 ]{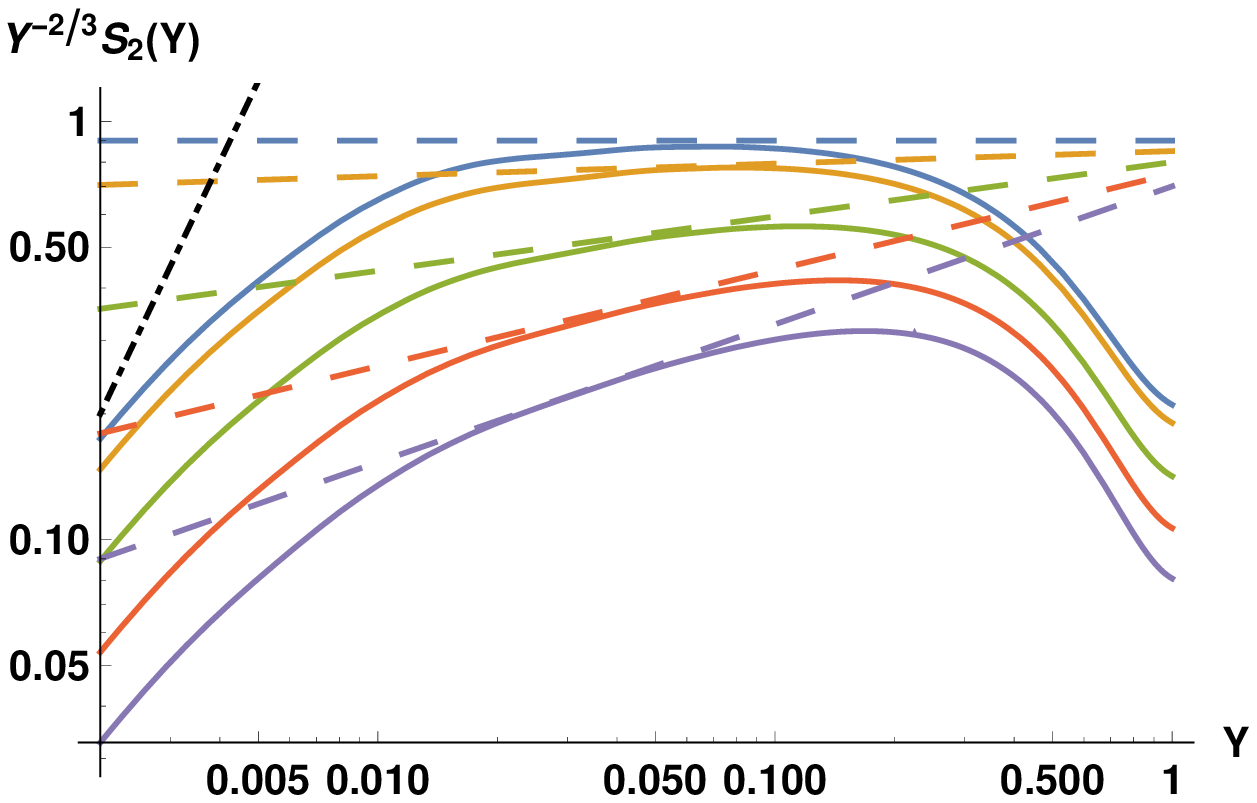} \\
    	\end{tabular}
    	\caption{  \label{F:2}   (a)Piecewise-linear model of the energy spectra $E(k)\propto k^{-m \Sb C}$ for $k<k_\times$ in the cascade-dominated interval and $E(k)\propto k^{-m\Sb F}, m\Sb F=3$ in the mutual friction-dominated and (b) structure functions $S_2(Y)$, computed using \Eq{1DFd}. The dashed lines of the matching colors correspond to $Y^{m\sb C -1}$. The black dot-dashed line  corresponds to $Y^2$.}
    \end{figure*}

	\begin{figure*}
		
		\begin{tabular}{ccc}
			(a)                        &                       (b)                       &                        (c)                        \\
		\includegraphics[scale=0.4   ]{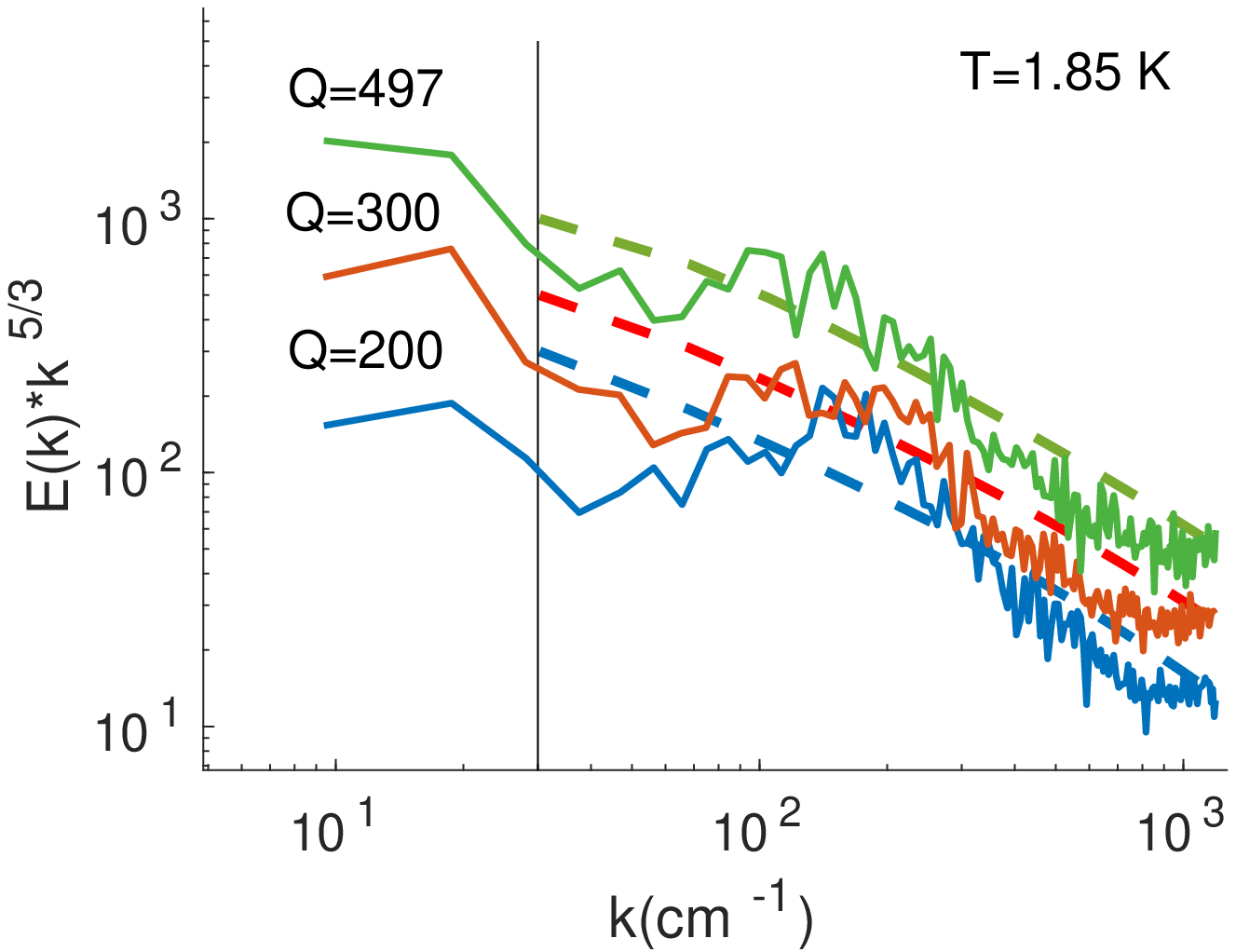}
		 & \includegraphics[scale=0.42 ]{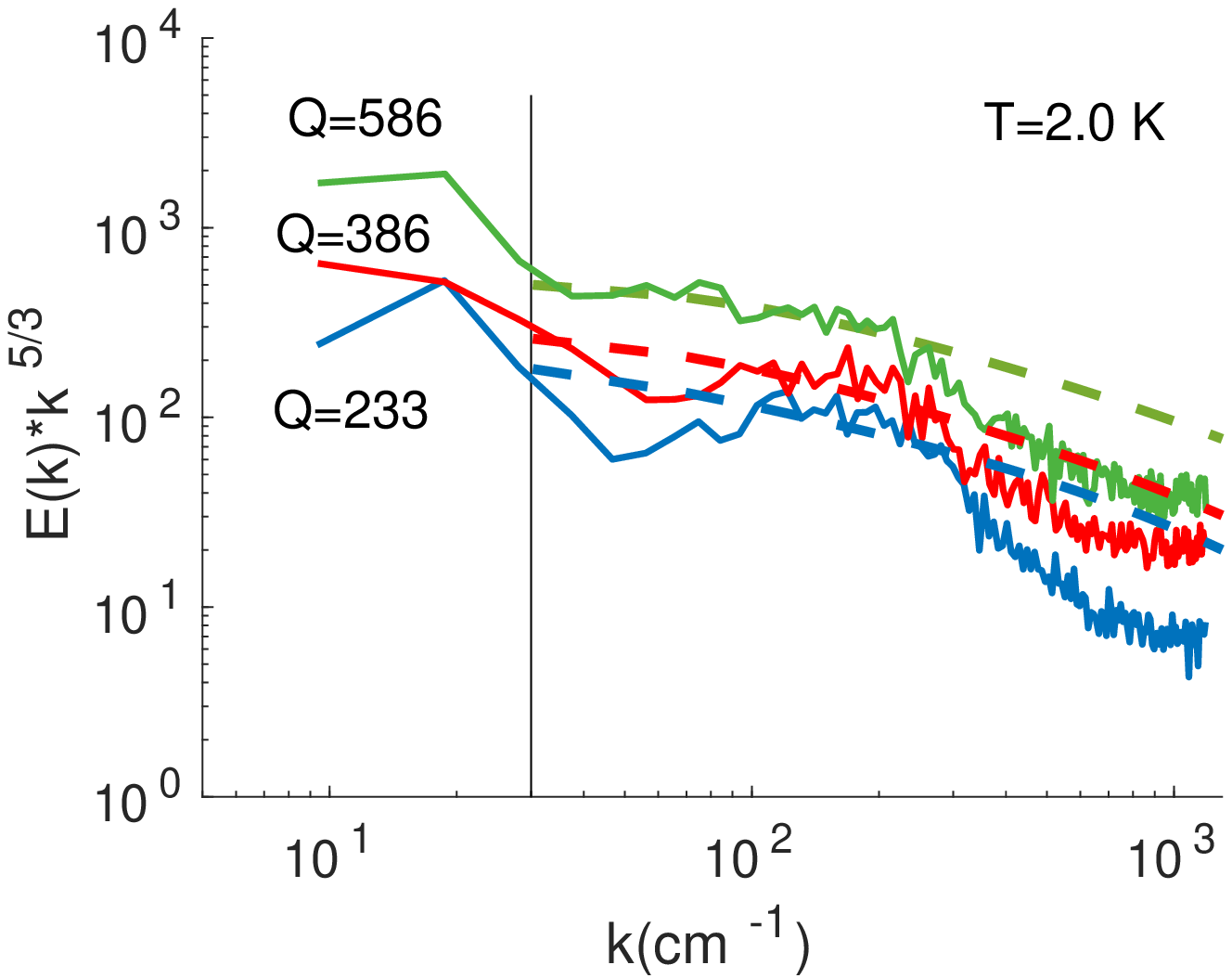}
	   & \includegraphics[scale=0.4   ]{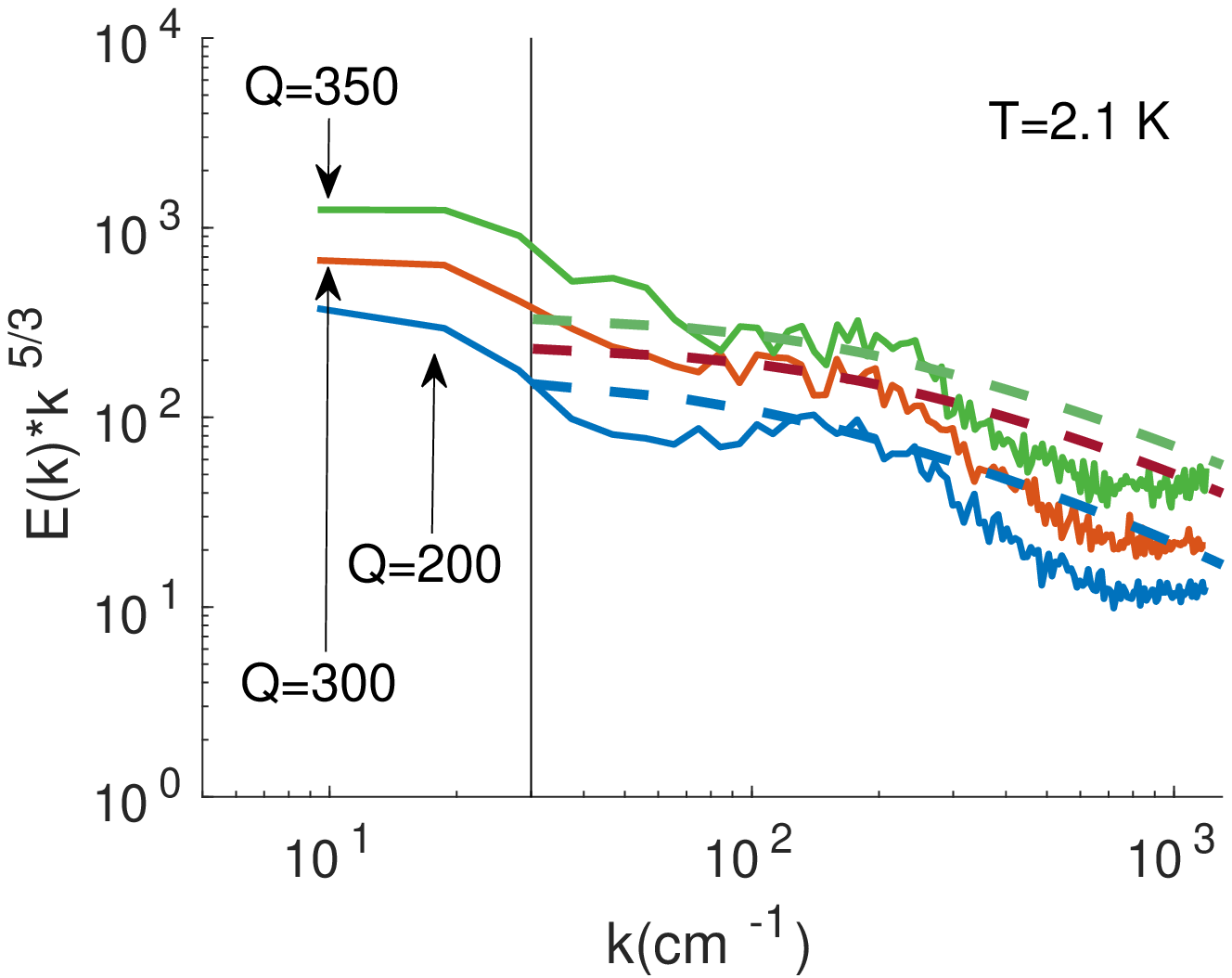}
		\end{tabular}
		\caption{  \label{F:3} Comparison of the experimental compensated energy spectra $k^{5/3}E(k)$ with the theoretical predictions\,\cite{LP-2018} (dashed lines)   for different temperatures $T\,$K and heat fluxes $Q$ (in mW/s$^2$).  Vertical solid lines denote the outer scale $k_0$.}.
	\end{figure*}

    \subsubsection{\label{sss:E}Energy spectra of counterflow turbulence}
    One-dimensional energy spectra, compensated by K41 scaling $k^{5/3}E(k)$, are shown  in \Figs{F:1}, middle column.   We can easily identify several  $k$-ranges with different  $k$-dependence of the spectra.  The small wavenumber ($k\lesssim k_0$)  behavior  is clearly different from the rest of the spectrum.  A relatively short part of the spectra is close to compensation by K41 while remaining steeper. The large-$k$ part of the spectra has a much larger slope, extending over all remaining interval of scales. We will now try to relate between these different types of behavior and various crossover wavenumbers introduced in \Sec{ss:est}.

   \paragraph{Energy-containing interval.}  The outer scale of turbulence,  $k_0\simeq 30$~cm$^-1$, was taken according to \Ref{WG-2017}, where it was estimated as $2\pi/r_0$ with $r_0\simeq 0.2\,$cm  close to    the maximum of the structure functions.
  The positions of $k_0$ are marked in \Fig{F:1}, middle and right columns,  by vertical black thin solid lines.  The range $k\lesssim k_0$ can be interpreted as an energy-containing interval, where   energy is pumped into the system due to instabilities of mean flow in the channel and   in which the most of  the flow energy is localized.   As we see, this value corresponds well to a boundary between the  large-scale behavior and the inertial-like scaling behavior of the spectra for large heat fluxes, while for low heat fluxes and low temperature, the  energy-containing interval seems  to extend  to higher wavenumbers.

    \paragraph{Cascade-dominated  interval.} Next characteristic scale, $k_\times$ (cf.  Table\,\ref{t:1}, Column \#9)   estimates the wavenumber  for which the decoupling function $D(k_\times)=\frac 12$.  For $k\lesssim k_\times$,  $D(k)>\frac 12$, and the energy dissipation by the mutual friction is relatively weak. In this $k$- range the main mechanism, responsible for the energy transfer over scales  is the Richardson-Kolmogorov energy cascade, similar to that in classical turbulence. Nevertheless the energy dissipation by the mutual friction cannot be fully ignored. Therefore the energy spectra in this range of scales are steeper than the K41 scaling, as is clearly seen in \Figs{F:1}, middle column.  All these motivate us to name the wavenumber range   $k_0\lesssim k \lesssim k_\times$ as \textit{cascade- dominated  interval}.

     It was suggested in \Ref{LP-2018} to characterize the apparent scaling of the otherwise not-scale-invariant spectra by calculation of a mean exponent over some interval of scales. The theoretical mean exponents over a first decade $\langle m\rangle_{10}$ were found to agree with the experimental exponents\cite{WG-2017} of the structure functions\cite{LP-2018}. We calculated the mean exponents over a $k$-range $k\in[30- 300]$ ( a decade in $k/k_0$) and collected them in Table\ref{t:1}, column \#14. These values are close to $n+1$,  where $n$ is apparent scaling exponent of $S_2(Y)\propto Y^n$, defined by \Eq{1DFd}.    This means that the idea to estimate $\< m\>_{10}$ via $n+1$ indeed works reasonably well.  The corresponding fits are shown in Fig.\ref{F:1} middle column as dashed colored lines. However, although the values of the mean exponents agree with the scaling of the structure functions, the actual scaling of the spectra is different.

   To estimate  the  apparent scaling exponents  $m \Sb C$ of the energy spectra $E(k)\propto k^{-m\Sb C}$ in the cascade-dominated  interval,  we plot the  experimental spectra, compensated by  $ k^{m\Sb C}$  and  choose  the value of $m \Sb C$ so as to maximize the  $k$-range where  $ k^{m\Sb C} E(k)\approx$ const. The resulting plots are shown in \Fig{F:1}, right  column, where the crossover scales $k_\times$ are shown by vertical dashed colored lines. It is remarkable that, except for the low heat fluxes at $T=1.65$ and $1.85$\,K, the cross-over between different scaling regimes of the energy spectra coincide well with $k_\times$.

      \paragraph{Mutual-friction  dominated interval.}

      For $k\gtrsim k_\times$,  where the decoupling function is small and mutual friction becomes important in the energy balance\,\cite{LP-2018},   the  slope of the energy spectra increases significantly  from $m\Sb C \approx 2.0 \pm 0.2$ to  $m\Sb F\approx  2.9\pm 0.1$. The transition between two types of behavior is not sharp, especially for low heat fluxes,   but clearly visible.

       The  power-law-like  behavior $E(k)\propto k^{-m\Sb F}$  qualitatively differs from the exponential decay of $E(k)$, typical for the viscous interval of $k$ in the classical hydrodynamic turbulence. Therefore, we consider this behavior as an evidence that for  $k\gtrsim k_\times$ the main mechanism of the energy dissipation is the mutual friction.   Upper limits of the mutual-friction  dominated interval $k_*$, are shown in \Fig{F:1}, right column, by vertical dot-dashed lines of the corresponding colors. As a rule, the values of $k_*$ are about or above the largest available values of $k$. This means that the viscous interval of wavenumbers is beyond   our spectral resolution. The corresponding $Y_*$ are shown in the  \Fig{F:1}, left column, and are mostly smaller than the implied boundary between the inertial and viscous behavior\cite{WG-2017} (i.e., black vertical thin solid lines). Therefore,  only at the smallest scales, the viscous dissipation becomes important, but it is still not dominant as we show below.

     \subsubsection{\label{sss:viscosity}More about connection between  $S_2(Y)$   and $E(k)$}
   To clarify the relation between the energy spectra in a finite $k$-range and the structure functions, we plot in
Fig.\,\ref{F:2} (a) a piecewise-linear model  of the energy spectra, consisting of $E(k)=k^{-m\Sb C}$ in the  cascade dominated interval $k_0 <k<k_\times$, continuously connected with the $E(k)\propto k^{-m\Sb F}$ part in the mutual-friction dominated interval $k_\times < k < k_*$.   We used the typical  values of $m\Sb C$,   and  the same values of $m\Sb F=3.0$, $k_\times=200\,$cm$^{-1}$, and $k_*=1300$ for all spectra. For simplicity, we adopt for  the energy containing interval the same behavior $E(k)=k^{-m\Sb C}$ as in the  cascade dominated interval. The structure functions, computed using \Eq{1DFd}, are shown in \Fig{F:2} (b) together with the  expected slope  $S_2(Y)\propto Y ^{n}$  with $n=m\Sb C-1$, shown by the dashed lines. As we see, the actual range of scales, over which the original scaling is recovered, is  very narrow.

 The slope $S_2(Y)\propto Y ^2$, typical for  viscous exponential decay of  $E(k)$ in the classical hydrodynamic turbulence, and
    $E(k)\propto k^{-3}$, typical for the mutual friction dominated interval in counterflow turbulence, are shown by black dot-dashed line.
    As expected, for the finite scaling interval of a modest extent, the resulting $Y$-dependence of   $S_2(Y)$ demonstrates very smooth transition between these regimes  and does not reach the genuine  asymptotic behavior  $S_2(Y)\propto Y^2$.
    	
\begin{figure*}
	\begin{tabular}{ccc}
		(a)                        &                       (b)                       &                        (c)                        \\
		\includegraphics[scale=0.4   ]{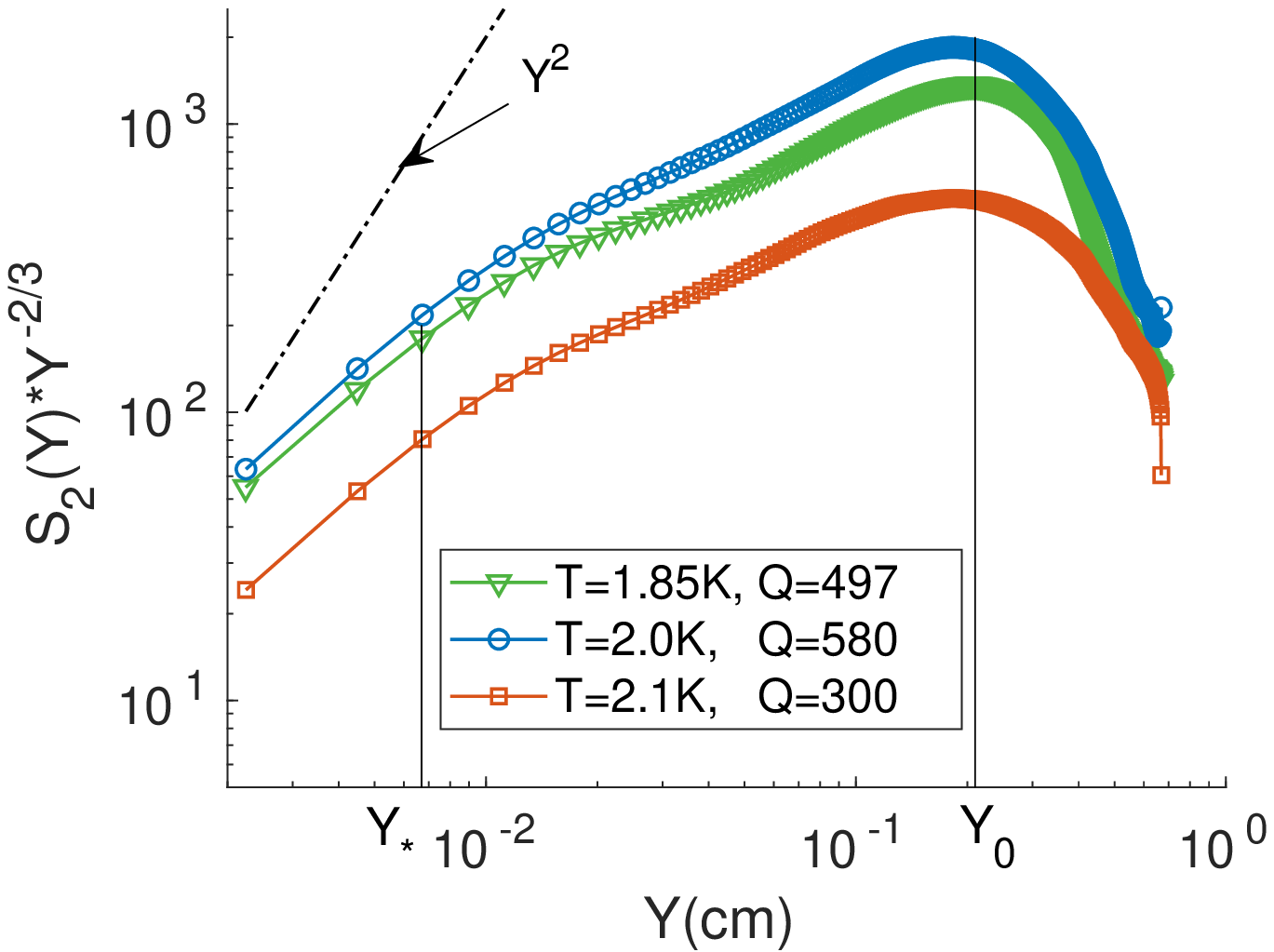} &
		\includegraphics[scale=0.4  ]{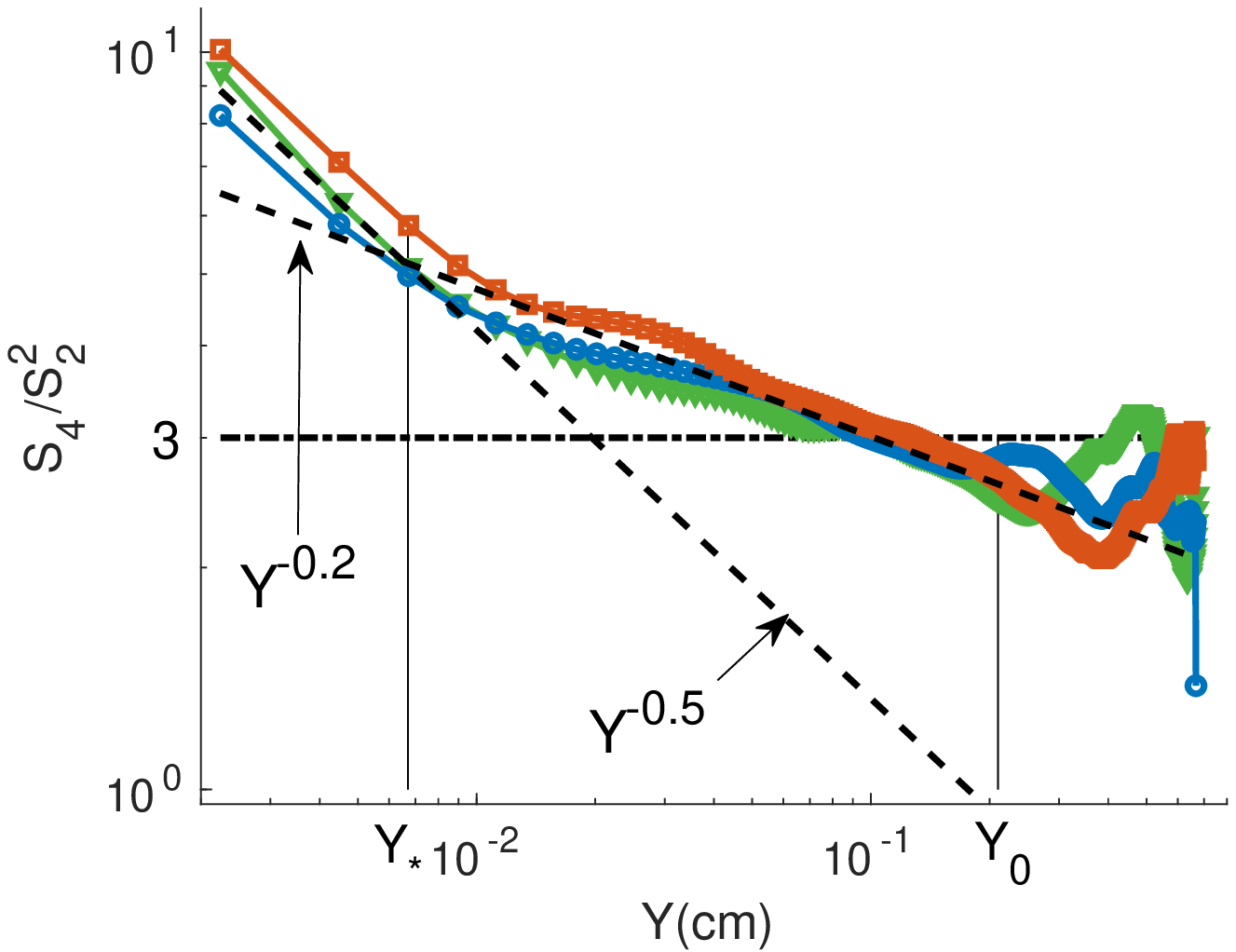} &
		\includegraphics[scale=0.4    ]{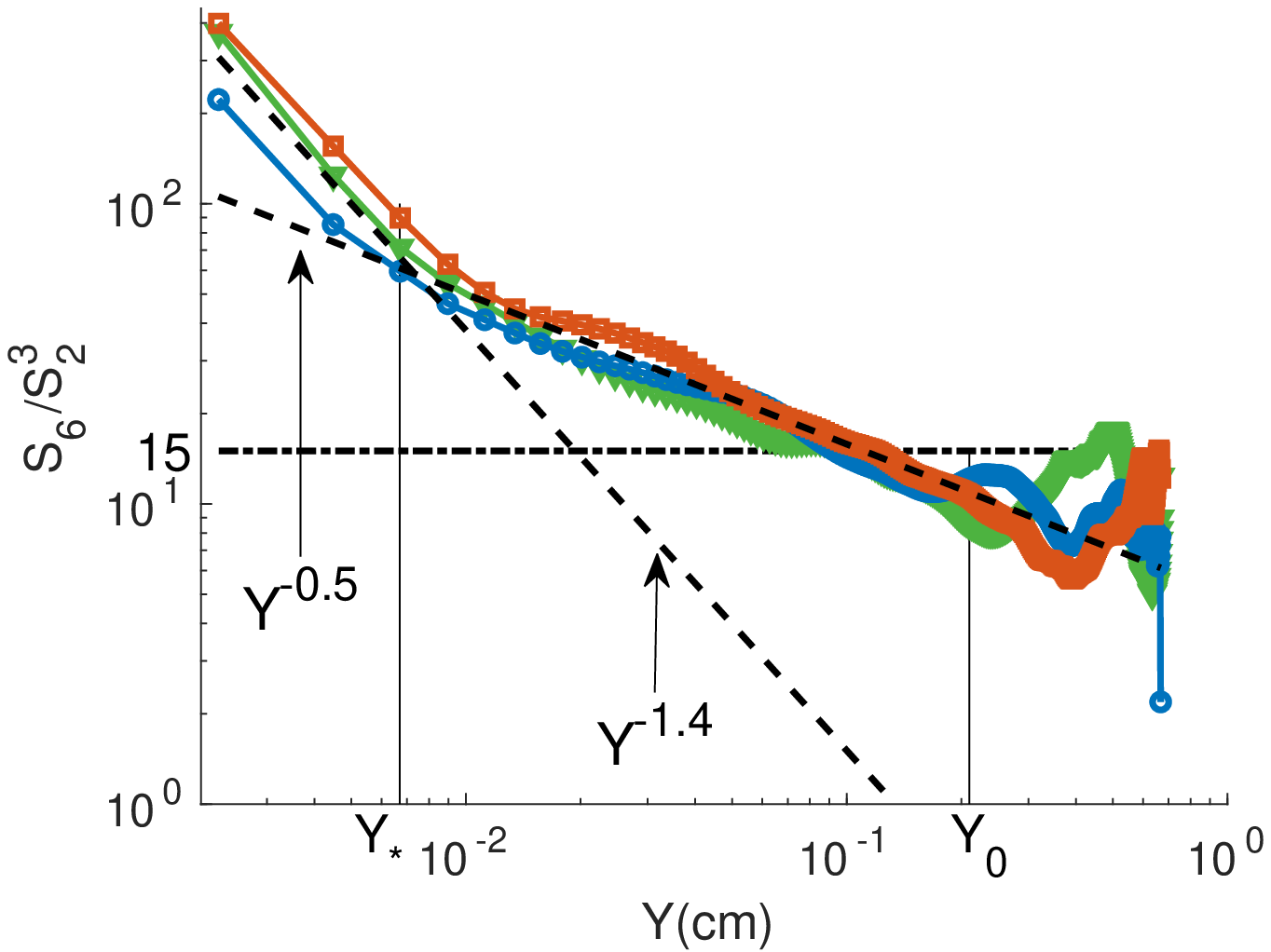}
	\end{tabular}
	\caption{  \label{F:4} Second- and higher-order statistics at different temperatures.  The flow parameters are shown in the legend of panel~(a). Panel (a): the  structure functions $S_2(Y)$.  Dot-dashed black line denotes the asymptotic viscous  behavior for $S_2(Y)\propto Y^2$.    Panel  (b):  Flatness $F_4(Y)=S_4(Y)/S_2^2(Y)$. Panel (c):   Hyper-flatness $F_6(Y)=S_6(Y)/S_2^3(Y)$. The horizontal dot-dashed lines in panels (b) and (c) denote the Gaussian values of $F_4(Y)$ and $F_6(Y)$.  The  approximate $Y$-dependencies of the flatness and hyperflatness for  $Y \gtrsim Y_*$  are  shown by dashed lines, marked $ Y^{-0.2}$ and $Y^{-0.5}$, respectively.   These  values are slightly larger than typical for the classical hydrodynamic turbulence, i.e., $\delta^{-0.14}$ and $\delta^{-0.38}$ in \Ref{zetas-2017}. However for $Y \lesssim Y_*$, the effective slopes of $F_4(Y)$ and $F_6(Y)$ strongly  increase and become much larger than the classical values, see the dashed lines, marked as $Y^{-0.5}$ and $Y^{-1.4}$, respectively. The vertical thin lines in all panels denote positions of the outer scale $Y_0$ and the crossover scale $Y_*$.  }
\end{figure*}

  \subsection{\label{ss:comp} Comparison of analytically predicted and experimental energy spectra }

It is instructive  to compare directly the experimental spectra with the spectra predicted by theory\cite{LP-2018}.
Remind that the theory does not describe the largest scales motion in the energy-containing interval (for $k<k_0$) and has the energy influx for $k=k_0$ (or the  boundary condition $E_0=E\sp{th}(k_0)$) as an external parameter of the theory.

The theory was developed for an idealized situation of fully-developed, space homogeneous turbulence. Naturally, the real physical situation in the  experiments (i.e., wall-bounded, space- inhomogeneous channel flow for relatively low Reynolds numbers) is more complicated than assumed by the theory. Therefore, the comparison is meaningful only for the experiment with relevant flow conditions. In our case, we may take the Re$\sb n>100$ as a tentative criterion for the well-developed turbulence in the channel. This leaves out the low temperature data ($T=1.65\,$K), as well as lowest heat fluxes for $T=1.85$ and $2.0\,$K. However we keep for completeness all the data for $T=1.85\,$K and $2.0\,$K.

In \Figs{F:3} we compare the experimental (K41-compensated) energy spectra for $T=1.85\,$K, $2.0\,$K and $2.1\,$K (plotted as solid colored lines) with the predicted  energy spectra (denoted by dashed lines of the same color), calculated  for  the same temperatures and the same heat fluxes. The theoretical spectra were made dimensional and $E\sp{th}(k_0)$ was taken to ensure overlap in  the cascade-dominated $k$-range. For high heat fluxes, these values agree well with  $E\sp{exp}(k_0)$.

The immediate observation is a qualitative agreement between the experimental and theoretical spectra over large range of wavenumbers, covering most of the cascade-dominated range. The deviations are mostly limited to the spectra with the lowest heat fluxes for $T=2.0\,$K and $T=2.1\,$K. At $T=1.85\,$K, the agreement is recovered for wavenumbers larger than $k_0$, which may indicate that the turbulence in these experiments is not yet fully developed and the outer scale is smaller that expected.

As mentioned above, the overall suppression of the spectra compared to the classical behavior is well captured by  the mean scaling exponents $\langle m\rangle_{10}$, cf. Table\ref{t:1}, column \#14.

On the other hand, the theory does not describe the sharp drop of the spectra in the mutual-friction dominated $k$-range,  demonstrating only smooth decrease of the current slope $m(k)$ of the spectra for large $k$.   A possible reason is that the theory\,\cite{LP-2018} does not take into account the energy exchange between components, that is most efficient in this range of scale. Other flow conditions, which are not accounted for by the theory, may contribute to this discrepancy.  Also the experimental data in the high $k$ regime may not be very reliable indeed. This is because the corresponding separation scale is comparable or even smaller than the width of the tracer line, which leads to large uncertainty. \\~

  \subsection{\label{ss:flat}  Flatness,   hyper-flatness  and  intermittency}
  To analyze higher order statistics and possible intermittency effects, we select one example at each temperature, having similar $Y_*$, and plot in  \Fig{F:4} the structure functions  $Y^{-2/3}S_2(Y)$, the  flatness $F_4(Y) =S_4(Y)/S_2^2(Y)$, and the  hyper-flatness $F_6(Y) =S_6(Y)/S_2^3(Y)$.

 In all panels, we mark the positions of the outer scale of turbulence $Y_0$ and the crossover scale $Y_*$.  As is clearly seen in  \Fig{F:4}a, the asymptotic behavior $Y^2$ is not  reached with our spatial resolution. However, $Y_*$ delineates different types of behavior of the $S_2(Y)$. These different regimes are even better exhibited by flatness and hyperflatness in \Figs{F:4}(b) and (c).

 Remind that for Gaussian statistics $F_4=3$ and $F_6=15$, shown in  \Figs{F:4} (b) and (c) as horizontal dashed lines.  Clearly,  for large scales  $Y  \gtrsim Y_0\,$,  $F_4(Y)$ and  $F_6(Y)$   are close to the  Gaussian values, indicating that the  statistics of the  turbulent velocity field in the energy-containing interval  is indeed close to Gaussian. This is a common property of classical hydrodynamic turbulence, independent of the way of its excitation.

 In a wide  interval of scales $Y_*\lesssim Y \lesssim Y_0$, covering scales corresponding to both the cascade-dominated and mutual-friction dominated spectral ranges, both $F_4(Y)$ (panel(b)) and $F_6(Y)$  (panel(c)) have a power-law-like behavior $F_n(Y)\propto Y ^{-x^{(1)}_n}$, with  exponents $x_4\sp{(1)}\simeq 0.20\pm 0.02$ and $x_6\sp{(1)}\simeq 0.5\pm 0.03$. To compare these exponents with their counterparts   $x_n\Sp{HT}$ in  the classical hydrodynamic turbulence, remind that  $x_n\Sp{HT}=\zeta_n\Sp{HT}- n\zeta _2\Sp{HT}/2$, where $\zeta_n\Sp{HT}$ is the scaling exponent of the $n$-order structure function in classical hydrodynamic turbulence.  With the most recent experimental values\,\cite{zetas-2017}   $\zeta_2\Sp{HT}\approx 0.72$, $\zeta_4\Sp{HT}\approx 1.30$ and $\zeta_6\Sp{HT}\approx 1.78$,  this gives  $x _4\Sp{HT}\approx 0.14$ and $x _6\Sp{HT}\approx 0.38$.  We conclude that  the values $x_{4,6}^{(1)}$ are moderately,  but distinctly larger than  $\zeta_n\Sp{HT}$.
 Notably, the structure functions and higher-order statistics is not sensitive to the peculiarities of the energy spectra, in particular to the existence of two significantly different scaling ranges.

However, at smaller scales $Y \lesssim Y_*$, the effective slopes of $F_4(Y)$ and $F_6 (Y)$ increase  dramatically. The estimates, shown in \Figs{F:4} (b) and (c), give $x_4\sp{(2)}\simeq 0.5\pm 0.1$ and $x_6\sp{(2)}\simeq 1.4\pm 0.1$
at these scales, which correspond to the dissipative range  with mixed contributions of the mutual-friction and viscous dissipations. The statistics become very intermittent. The fact that we do not observe saturation of  $F_4(Y)$ and $F_6 (Y)$, typical for the viscous range in the classical turbulence,  supports our conjecture that the viscous dissipation-dominant range is beyond our resolution.

 \section{\label{s:sum} Conclusion}
 	In this paper, we report a detailed analysis of  the energy spectra, second- and high-order  structure functions of velocity differences in the  superfluid $^4$He counterflow turbulence, measured in a wide range of temperatures and heat fluxes.  In particular,  we discover two ranges  of wavenumbers $k_y$ with very different apparent exponents of the  one-dimentional energy spectra  in the  cascade-dominated (for relatively small $k_y$)  and  the  mutual friction-dominated  subintervals, respectively. The general behavior  of the experimental spectra $E_{xz}(k_y)$ in the cascade-dominated range agrees well with the predicted   energy spectra\, in \Ref{LP-2018}.
 	
 		The analysis of the statistics of the high-order  structure functions shows that in the energy-containing interval the statistics of counterflow turbulence is close to Gaussian, similar to  the classical hydrodynamic turbulence. In the  cascade- and mutual friction-dominated intervals we found some modest  enhancement of intermittency with respect of its level in classical turbulence. However,  at  small  scales (but not yet viscous scales), the  intermittency becomes much stronger than that  in the classical turbulence.

 In conclusion,  we should remind that the theory, developed in \Ref{LP-2018}, does not describe the experimental observations reported here in all details. Besides the obvious reason of space inhomogeneity, especially important for modest available Reynolds  numbers Re$\sb n$, there is one more possible  reason for some disagreement even in the Re$\sb n\to \infty$ limit.   This is the anisotropy of statistics of counterflow turbulence. Although we do not have yet experimental information how strong the anisotropy of turbulent statistics is, this effect is definitely there due to presence of preferred $\B U\sb{ns}$ direction and  strong dependence of the cross-correlation function $E\sb{ns}(\B k)$  (between the normal fluid and superfluid velocity components) on the angle between $\B U\sb{ns}$ and $\B k$, predicted in  \Ref{decoupling}.  The study the effect of anisotropy on the statistics of counterflow turbulence is in our nearest agenda. Nevertheless, the reasonable agreement between our  observations   and the theory\,\cite{LP-2018}, is encouraging.  In particular,  the  crossover scales between different regimes, predicted by the theory using macroscopic parameters of the flow,  and clearly observed in the spectra and structure functions,   makes us believe   that what we know so far contains essential part  of the basic physics of the problem.

 	\acknowledgments
 S. B and W.G. acknowledge support from the National Science Foundation under Grant No. DMR-1807291. The
experiment was conducted at the National High Magnetic Field Laboratory, which is supported by
NSF Grant No. DMR-1644779 and the state of Florida. The authors would also like to acknowledge J. Gao for assistance in processing the experimental data.

\end{document}